\documentclass[journal]{IEEEtran}
\usepackage{cite,color}

\ifCLASSINFOpdf
   \usepackage[pdftex]{graphicx}
   \graphicspath{{../pdf/}{../jpeg/}}
   \DeclareGraphicsExtensions{.pdf,.jpeg,.png}
\else
   \usepackage[dvips]{graphicx}
   \graphicspath{{../eps/}}
   \DeclareGraphicsExtensions{.eps}
\fi

\usepackage{amsmath,amsfonts,amsthm,bm}
\usepackage{bbm,multicol,multirow,makecell}
\usepackage{array}
\usepackage{algorithm,algorithmic,float}

\ifCLASSOPTIONcompsoc
  \usepackage[caption=false,font=normalsize,labelfont=sf,textfont=sf]{subfig}
\else
  \usepackage[caption=false,font=footnotesize]{subfig}
\fi

\ifCLASSOPTIONcaptionsoff
  \usepackage[nomarkers]{endfloat}
 \let\MYoriglatexcaption\caption
 \renewcommand{\caption}[2][\relax]{\MYoriglatexcaption[#2]{#2}}
\fi
\usepackage{url}
\newtheorem{proposition}{Proposition}
\newtheorem{theorem}{Theorem}
\newtheorem{lemma}{Lemma}
\newtheorem{remark}{Remark}
\newtheorem{definition}{Definition}

\begin{document}

\title{Graph-Embedded Multi-Agent Learning for Smart Reconfigurable THz MIMO-NOMA Networks}

\author{Xiaoxia~Xu,~\IEEEmembership{}
        Qimei~Chen,~\IEEEmembership{Member,~IEEE,}
        Xidong~Mu,~\IEEEmembership{Graduate~Student~Member,~IEEE,}
        Yuanwei~Liu,~\IEEEmembership{Senior~Member,~IEEE,}
        and~Hao~Jiang,~\IEEEmembership{Member,~IEEE}
\thanks{
X. Xu, Q. Chen, and H. Jiang are with the School of Electronic Information, Wuhan University, Wuhan 430072, China (e-mail: \{xiaoxiaxu, chenqimei, jh\}@whu.edu.cn).}
\thanks{X. Mu is with School of Artificial Intelligence, Beijing University of Posts and Telecommunications, Beijing, China. (email: muxidong@bupt.edu.cn).}
\thanks{Y. Liu is with the School of Electronic Engineering and Computer Science, Queen Mary University of London, London E1 4NS, UK. (email:yuanwei.liu@qmul.ac.uk).}
}

%
%

\markboth{}%
{Shell \MakeLowercase{\textit{et al.}}: Bare Demo of IEEEtran.cls for IEEE Journals}
%



\maketitle

\begin{abstract}
With the accelerated development of immersive applications and the explosive increment of internet-of-things (IoT) terminals, 6G would introduce terahertz (THz) massive multiple-input multiple-output non-orthogonal multiple access (MIMO-NOMA) technologies to meet the ultra-high-speed data rate and massive connectivity requirements.
Nevertheless, the unreliability of THz transmissions and the extreme heterogeneity of device requirements pose critical challenges for practical applications.
To address these challenges, we propose a novel smart reconfigurable THz MIMO-NOMA framework, which can realize customizable and intelligent communications by flexibly and coordinately reconfiguring hybrid beams through the cooperation between access points (APs) and reconfigurable intelligent surfaces (RISs).
The optimization problem is formulated as a decentralized partially-observable Markov decision process (Dec-POMDP) to maximize the network energy efficiency, while guaranteeing the diversified users' performance, via a joint RIS element selection, coordinated discrete phase-shift control, and power allocation strategy.
To solve the above non-convex, strongly coupled, and highly complex mixed integer nonlinear programming (MINLP) problem, we propose a novel multi-agent deep reinforcement learning (MADRL) algorithm, namely \emph{graph-embedded value-decomposition actor-critic (GE-VDAC)}, that embeds the interaction information of agents, and learns a locally optimal solution through a distributed policy.
Numerical results demonstrate that the proposed algorithm achieves highly customized communications and outperforms traditional MADRL algorithms.

\end{abstract}


\IEEEpeerreviewmaketitle

\section{Introduction}
\IEEEPARstart{T}{he} upcoming 6G era confronts a variety of emerging applications, involving immersive applications such as ultra-high definition (UHD) video and virtual reality/augmented reality (VR/AR), as well as Internet of Things (IoT) applications like wearable devices and smart homes \cite{6GUseCase_Giordani_2020}.
To meet the unprecedented challenges raised by ultra wide-band communications and massive IoT connectivity, terahertz (THz) massive multiple-input-multiple-output non-orthogonal multiple access (MIMO-NOMA) has become an essential technology for 6G, which can provide $10$ Gbps-order ultra-fast transmission speed and support millions of connections.
Generally, THz MIMO-NOMA systems utilize the large-scale antenna array with hybrid beamforming structure \cite{DynamicSubarray_Yan_2020}, which can compensate the severe fading over high-frequency THz bands, and reduce the hardware complexity and power consumption.
In addition, assisted with the MIMO-NOMA technology \cite{MIMONOMA_Liu_2017,MIMONOMA_Liu_2018}, highly spatial-correlated users can be grouped into one cluster and supported by a single radio frequency (RF) chain, which can significantly improve spectral efficiency and connective density \cite{MIMONOMA_Zeng_2017}.
To employ the massive MIMO-NOMA technology for THz communications, the authors in \cite{THzMIMONOMA_Zhang_2020} proposed an energy-efficient user clustering, hybrid precoding, and power optimization scheme, where the blockage probability and unreliability of line-of-sight (LoS) links have been ignored.
Nevertheless, due to the high obscuration susceptibility, the application of THz MIMO-NOMA network may suffer from serious transmission unreliability and intermittency resulting from either wall blockage or human-body blockage effect \cite{Blocakge_Wu_2021}, which may significantly degrade the experiences of unreliability-sensitive 6G immersive applications.

Fortunately, the newly-emerged reconfigurable intelligent surface (RIS) technology is regarded as a promising way to overcome the shortcomings in THz MIMO-NOMA communications \cite{SmartWireless_Gong_2020,SmartRadio_Di_2020,RIS_Liu_2021}. Specifically, RISs can dynamically transform and reshape spatial beams, and thus construct virtual LoS links between transmitters and receivers to avoid blockage \cite{ReliabVR_Chaccour_2020}.
Meanwhile, a smart radio environment can also be created to achieve significant spectrum/energy efficiency improvement and flexible scheduling \cite{RISNOMAinterplay_Liu_2020,RISNOMA_Mu_2021}.
Given the aforementioned benefits, increasingly research efforts have been devoted to the RIS-aided MIMO-NOMA networks operating in low frequencies.
In \cite{MISONOMARIS_Li_2019}, the authors proposed a joint passive and active beamforming method for RIS-aided multiple input single output (MISO)-NOMA network, which obtained a locally optimal solution based on a second-order cone programming (SOCP)-alternating direction method of multipliers (ADMM) algorithm.
The authors in \cite{RISNOMABF_Mu_2020} further studied the joint passive and active beamforming in RIS-aided MISO-NOMA networks under both ideal and non-ideal RIS assumptions.
Furthermore, the authors in \cite{RISNOMA_Liu_2020} proposed joint deployment and beamforming frameworks for RIS-aided MIMO-NOMA networks based on deep reinforcement learning.
However, the existing methods are inapplicable to 6G THz MIMO-NOMA communications:
1) The existing MIMO-NOMA mechanisms are incapable to deal with the extremely heterogenous quality-of-service (QoS) requirements of 6G users.
2) Compared with low-frequency MIMO communications, THz MIMO communications face a more prominent transmission unreliability problem.
3) Since THz MIMO-NOMA networks usually have high-dimensional spatial channel information, existing centralized and iterative optimization algorithms usually lead to unacceptable high complexity and information exchange overhead to schedule the complicated THz MIMO-NOMA scenarios.

Therefore, we aim to propose a novel smart reconfigurable MIMO-NOMA THz framework that can realize customizable and intelligent indoor communications with high energy efficiency and low complexity based on a machine learning mechanism in this work.
Here, we consider two types of heterogeneous users, namely IoT users and super-fast-experience (SE) users.
Specifically, SE users, like VR/AR and UHD video, require ultra-high-speed and reliable immersive communication experiences, while the densely connected IoT users, like smart cleaners and smart watches, tolerate sporadic and unreliability-tolerant traffic transmission.
Different from the conventional systems, we introduce a low-complexity and decentralized learning-based framework that can jointly design the user clustering, NOMA decoding, and hybrid beam reconfiguration schemes in a cooperative setting with multiple APs and RISs:
1) By adaptively aligning users' equivalent spatial channels as well as customizing NOMA decoding orders based on the QoS requirements, the intra-cluster interference suffered by SE users can be completely eliminated.
2) We adjust the highly-directional hybrid beams through the cooperation among APs and RISs, which can ensure tailored spatial data channels and mitigate both inter-cluster and inter-AP interference via active hybrid beamforming and coordinated passive beamforming.
3) To overcome the non-ideal discrete phase-shift control, we exploit a dynamic RIS element selection structure for the hybrid beam reconfiguration, which can flexibly refrain unfavorable and negative reflections via an element-wise ON/OFF control to enhance energy efficiency.
Overall, the proposed framework can realize customizable hybrid spatial and power-domain multiplexing, as well as improving the multi-domain resource utilization.

Based on the above framework, we propose a long-term joint RIS element selection, coordinated discrete phase-shift control, and power allocation learning strategy. The objective function is formulated to maximize the system energy efficiency as well as satisfying users' data rate and reliability, which is an NP-hard mixed-integer nonlinear programming (MINLP) problem.
To efficiently solve the non-convex, strongly coupled, and highly complex MINLP problem online, we transfer it into a decentralized partially observable Markov decision process (Dec-POMDP).
Thereafter, we introduce a novel cooperative multi-agent reinforcement learning (MADRL) method, namely \textit{graph-embedded value-decomposition actor-critic (GE-VDAC)}, to efficiently coordinate multi-AP and multi-RIS in a distributed manner. The proposed \emph{GE-VDAC} algorithm can not only improve generalization ability of multi-agent learning, but also reduce information exchange overhead.

The main contributions of this work can be summarized as follows.
\begin{itemize}
\item We propose a novel smart reconfigurable THz MIMO-NOMA framework, which can realize highly customizable and intelligent communications to support ultra-wide bands and ultra-dense connections.
    The hybrid spatial beams are smartly and cooperatively reconfigured through multi-AP and multi-RIS coordinations, where dynamic RIS element selection, on-demand data enhancement, flexible interference suppression, and efficient hybrid spatial and power-domain multiplexing are allowed.
\item The long-term joint RIS element selection, coordinated discrete phase-shift control, and power allocation optimization problem is formulated by a Dec-POMDP model. Under customized user clustering, NOMA decoding, and sub-connected hybrid beamforming schemes, the Dec-POMDP model can maximize the expected system energy efficiency while satisfy extremely heterogeneous data rate and reliability requirements for different users.
\item To efficiently solve the non-convex, strongly coupled, and highly complex MINLP problem online, we propose a novel distributed MADRL algorithm, namely \textit{GE-VDAC}, which learns the decomposed local policies by embedding the agents' interaction information into dimension-reduced and permutation-invariant features.
    We show that the proposed \textit{GE-VDAC} can not only converge to a locally optimal solution, but also achieve a better coordination and generalization with low information exchange overhead.
\item We present numerical results to verify the effectiveness of the proposed strategy. The proposed \textit{GE-VDAC} achieves higher system energy efficiency and faster learning speed than traditional MADRL algorithms. Moreover, both reliable and ultra-high-speed communications can be achieved by SE users despite the increment of connected users.
\end{itemize}

The rest of this paper is organized as follows. Section II describes the smart reconfigurable THz MIMO-NOMA network. Section III presents the formulated Dec-POMDP problem, and Section VI proposes the \emph{GE-VDAC} based distributed MADRL algorithm. Numerical results are presented in Section V before the conclusion in Section VI.

\emph{Notation:} We denote the imaginary unit by $j$, and represent vectors and matrices by lower and upper boldface symbols, respectively.
$\frac{\partial F}{ \partial x}$ denotes the first partial derivative of function $F$ with respect to $x$.
$\mathbb{E}\left[\cdot\right]$ represents the statistical expectation.
$|\cdot|$ and $\|\cdot\|$ denote the absolute value and the Euclidean norm, respectively.
Moreover, $\star$ means the permutation operation.
The main notations used throughout this paper is summarized in Table \ref{table:notation}.
\begin{table}[t]
	\begin{minipage}{1\linewidth}
		\centering
		\caption{Main Notations}
		\label{table:notation}
		\resizebox{1\textwidth}{!}{
			\begin{tabular}{l|c}
				\hline
				Notations & Definitions \\ \hline
				$N_{\mathrm{A}}$, $N_{\mathrm{R}}$                 & number of antennas and RF chains    \\
                $N_{\mathrm{sub}}$                                 & number of antennas in a subset \\
                $\mathbf{V}^{m},\mathbf{W}^{m}$                    & analog and digital beamformers at AP $m$  \\
                $K_{\mathrm{S}}$, $K_{\mathrm{U}}$	       & number of SE and IoT users associated to each AP  \\
				$K_n^{m}(t)$ ($\mathcal{K}_n^{m}(t)$)	           & user number (set) served by RF chain $n$ of AP $m$  \\
                $\bm{\Theta}^j(t)$                         & phase-shift matrix of RIS $j$ \\
                $\omega_{l}^{j}(t)$,  $\theta_{l}^{j}(t)$          & ON/OFF state and phase shift of element $l$ at RIS $j$ \\
                $p_{nk}^{m}(t)$, $P_{n}^{m}(t)$                    & power allocation of user $U_{nk}^{m}$ and cluster $n$ \\
                $\mathbf{h}_{nk}^{im}(t)$                          & equivalent channel from AP $i$ to user $U_{nk}^{m}$ \\
                $\widetilde{\mathbf{h}}_{nk}^{im}(t)$              & direct channel from AP $i$ to user $U_{nk}^{m}$ \\
                $\mathbf{f}_{nk}^{jm}(t)$                          & channel from RIS $j$ to user $U_{nk}^{m}$ \\
                $\mathbf{G}^{ij}(t)$                               & channel from AP $i$ to RIS $j$ \\
                $q_{nk}^{m}(t)$, $Y_{nk}^{m}(t)$                   & traffic queue and virtual queue of user $U_{nk}^{m}$ \\
                $\mathbf{\Omega}(t)$, $\mathbf{s}(t)$, $\mathbf{a}(t)$ &  joint observation, state, and action vectors\\
                $\mathbf{o}_{i}^{u}(t)$                            & vectorized node feature of the $u$-th type agent $i\in\mathcal{I}^{u}$ \\
                $\mathbf{d}_{ii'}(t)$                              & vectorized feature of edge $(i,i')$ \\
                $\mathbf{e}_{ii'}(t)$                              & embedded feature of edge $(i,i')$ \\
                $\mathbf{z}_{i}^{u}(t)$, $\widetilde{\mathbf{z}}_{i}^{u}(t)$   & hidden and local embedded states of agent $i\in\mathcal{I}^{u}$ \\
                \hline
			\end{tabular}
		}
	\end{minipage}
\hfill
\end{table}

\section{Smart Reconfigurable THz MIMO-NOMA Framework}
\begin{figure}
  \centering
  \includegraphics[width=0.5\textwidth]{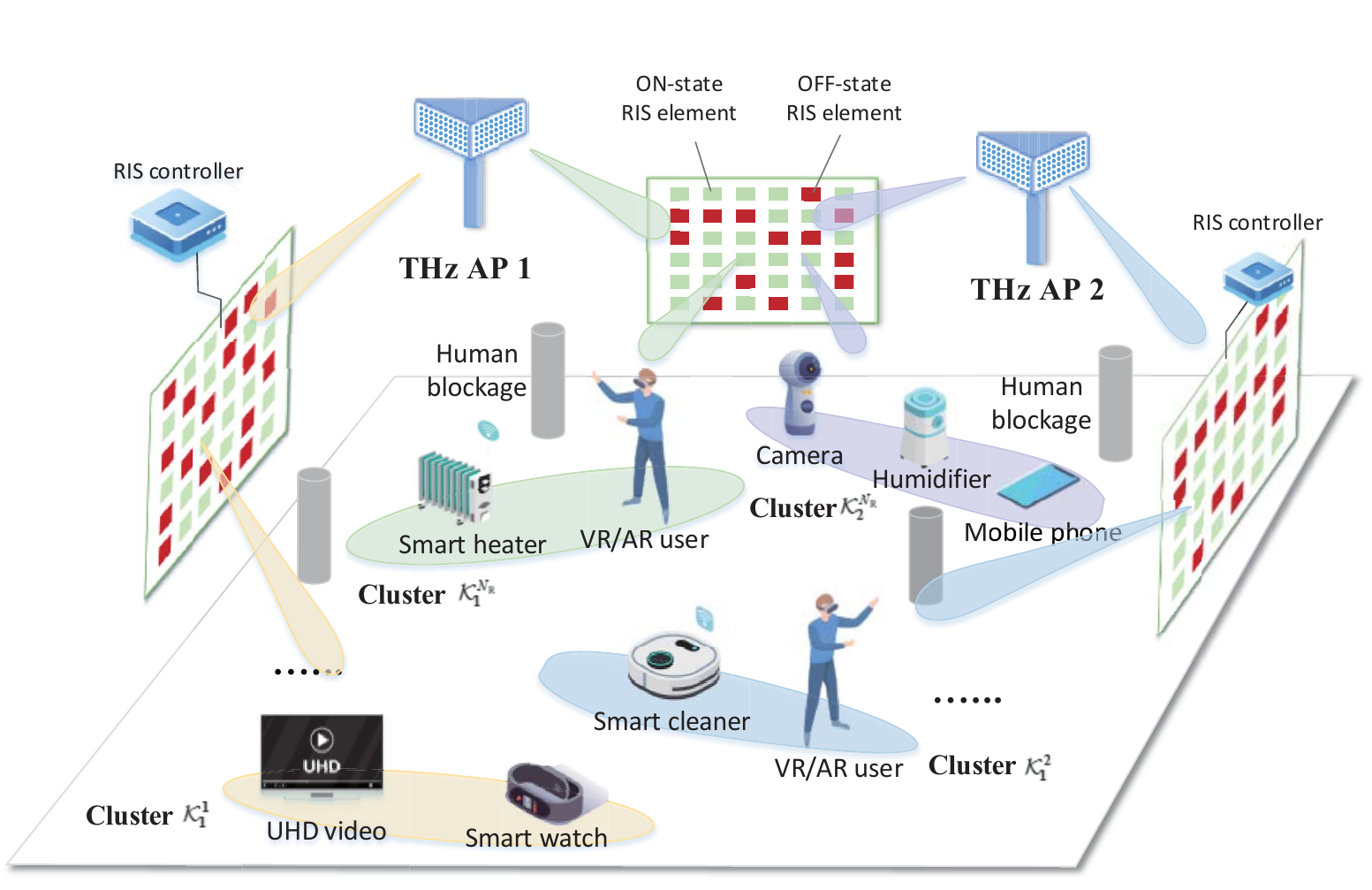}\\
  \caption{The proposed smart reconfigurable THz massive MIMO-NOMA network.}
  \label{fig_sysmodel}
\end{figure}
As shown in Fig. \ref{fig_sysmodel}, we consider an indoor downlink massive MIMO-NOMA THz network that serves densely distributed SE users and IoT users under multiple pre-installed THz APs and RISs.
Denote the set of $M$ APs and $J$ RISs as $\mathcal{M} = \{1,2,...,M\}$ and $\mathcal{J}= \{1,2,...,J\}$, respectively.
By coordinating and cooperating with neighboring APs and RISs, each AP $m$ serves a set $\mathcal{K}_{\mathrm{S}}^{m}$ of $K_{\mathrm{S}}^{m}$ SE users and a set $\mathcal{K}_{\mathrm{U}}^{m}$ of $K_{\mathrm{U}}^{m}$ IoT users, which have utterly diversified QoS requirements in terms of transmission data rate and reliability. For simplicity, we assume $K_{\mathrm{S}}^{1}=...=K_{\mathrm{S}}^{M}=K_{\mathrm{S}}$ and $K_{\mathrm{U}}^{1}=...=K_{\mathrm{U}}^{M}=K_{\mathrm{U}}$.
Here, we denote $\mathcal{K}^{m} = \mathcal{K}_{\mathrm{U}}^{m} \cup \mathcal{K}_{\mathrm{S}}^{m}$, $K^m=K_{\mathrm{U}}+K_{\mathrm{S}}$, and $\mathcal{K} =  \mathcal{K}^1 \cup \mathcal{K}^2 \cup ... \cup \mathcal{K}^M$.
To reduce hardware complexity and energy dissipation, we assume each user equips with a single antenna and the APs apply the sub-connected hybrid beamforming structure \cite{AdaptiveHP_Zhu_2016}.
Moreover, each AP is equipped with $N_{\mathrm{A}}$ antennas and $N_{\mathrm{R}}$ RF chains, where each RF chain connects to a subset of $N_{\mathrm{sub}}$ antennas via $N_{\mathrm{sub}}$ phase shifters with $N_{\mathrm{sub}} = N_{\mathrm{A}}/N_{\mathrm{R}}$.
Define the analog and digital beamforming matrixes at each AP $m$ as $\mathbf{V}^m\in \mathbb{C}^{N_{\mathrm{A}} \times N_{\mathrm{R}}}$ and $\mathbf{W}^m \in \mathbb{C}^{N_{\mathrm{R}} \times N_{\mathrm{R}}}$.
Denote the phase-shift matrix of each RIS $j$ by $\bm{\Theta}^j$.
Based on the MIMO-NOMA technology, the highly spatial-correlated SE and IoT users can be grouped into one cluster, which is served by a reconfigured beam transmitted from an AP antenna subarray that is connected with a RF chain.
Define $U_{nk}^{m}$ as user $k$ in cluster $n$ under AP $m$, and $\mathcal{K}_n^m$ be the set of $K_n^m$ users in cluster $n$ under AP $m$.
Furthermore, the whole system is divided into $T$ time slots, indexed by $\mathcal{T}=\{1,2,...,T\}$.

\begin{figure}
  \centering
  \includegraphics[width=0.42\textwidth]{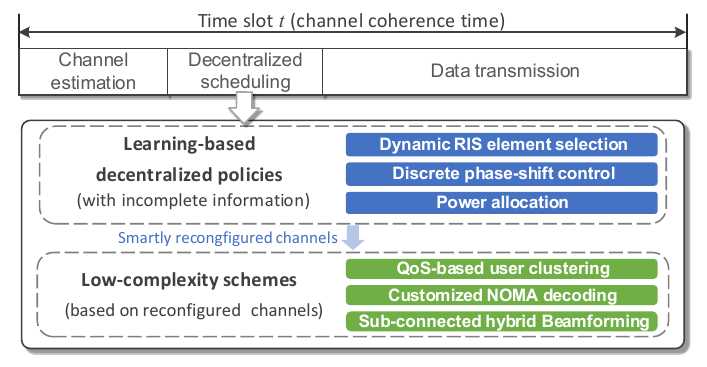}\\
  \caption{Process of the proposed learning-based mechanism.}
  \label{fig_procedure}
\end{figure}

In general, we aim to propose a learning-based mechanism for the smart reconfigurable THz
MIMO-NOMA network to jointly coordinate both multiple APs and RISs, which has the flow chart as shown in Fig. \ref{fig_procedure}.
Decentralized scheduling begins after channel estimation, which contains the process of learning-based decentralized policies and low-complexity schemes. In the learning-based decentralized policies, we can determine the dynamic RIS element selection, discrete phase-shift control, and power allocation with incomplete system information. Based on the smartly reconfigured channels, each AP can further obtain QoS-based user clustering, customized NOMA decoding order, and sub-connected hybrid beamformer using low-complexity schemes. Thereafter, data transmissions can be  performed.

Without loss of generality, we assume each RIS equips with a low-cost RF chain that can estimate the CSI based on the semi-passive RIS channel estimation methods \cite{RISCE_2021}. Specifically, at the beginning of each time slot, APs would send pilot signals to neighboring users to estimate the direct channels. Meanwhile, RISs would turn off their reflecting elements to sense the channels, where the indirect channels can be estimated by processing the received signals. Both direct and indirect channels can be estimated under compressed sensing or deep learning methods, which are out of the scope of this work.
For the sake of expression, we ignore the time slot index $t$ from Section \ref{Sec_RIS} to \ref{Sec_Power}.

\subsection{RIS Element Selection and Discrete Phase-Shift Control}\label{Sec_RIS}
We suppose that each RIS $j$ consists of $L$ reflecting elements, which is controlled by a software-defined RIS controller.
Since the continuous phase control is hard to realize in practice, a finite-resolution passive phase shifter is utilized for each reflecting element.
To save energy consumption as well as overcoming the non-ideal reflecting effect due to the discrete phase-shift control, we propose to leverage a dynamic RIS element selection structure during transmissions\footnote{Note that dynamic RIS element selection requires a complex design for the RIS array, which might be realized in the near future. In this work, we assume the reflecting elements in RISs can be dynamically turned ON/OFF by controlling the PIN diodes.}.
By flexibly controlling the ON/OFF states and the phase shifts of passive phase shifters, the dynamic selection structure can prevent unfavorable reflections and achieve higher energy efficiency.


Based on the dynamic selection structure, the phase-shift matrix $\bm{\Theta}^j$ on RIS $j$ can be given by
\begin{equation}
\bm{\Theta}^{j} = \mathrm{diag}\left(\omega_{1}^{j}e^{j \theta_{1}^{j}},...,\omega_{L}^{j}e^{j \theta_{L}^{j}}\right),
\end{equation}
where $\omega_{l}^{j}$ denotes the ON/OFF state of the $l$-th RIS element, given by
\begin{equation}
\omega_{l}^{j}\!=\!
\begin{cases}
1, & \text{if reflecting element $l$ at RIS $j$ is turned ON,}\\
 0, & \text{otherwise.}
 \end{cases}
 \end{equation}
The discrete phase shift of a selected element $l$ at RIS $j$ is determined by an integer $\beta_{l}^{j}\in \{0,1,...,2^{B^{\mathrm{R}}-1}\}$ and the RIS resolution bit $B^{\mathrm{R}}$, which can be written as
\begin{equation}
    \theta_{l}^{j} \in \mathcal{F} = \left\{ 2^{1-b}\pi \beta_{l}^{j} \big| \beta_{l}^{j} \in \{0,1,...,2^{B^{\mathrm{R}}-1}\} \right\}.
\end{equation}

\subsection{Channel Model}
The equivalent channel vector from AP $i\in\mathcal{M}$ to user $U_{nk}^{m}$ via multiple RISs is defined as $\mathbf{h}_{nk}^{im} = [\mathbf{h}_{nk}^{im1}, ..., \mathbf{h}_{nk}^{imN_{\mathrm{R}}}] \in \mathbb{C}^{1\times N_{\mathrm{A}}}$, where $\mathbf{h}_{nk}^{ims}\in\mathbb{C}^{1\times N_{\mathrm{sub}}}$ is the equivalent channel from subarray $s$ at AP $i$ to user $U_{nk}^{m}$.
Here, $\mathbf{h}_{nk}^{im}$ is given by
\begin{equation}
\mathbf{h}_{nk}^{im} = \underset{\text{AP-user}}{\underbrace{\widetilde{\mathbf{h}}_{nk}^{im}}}
+\sum_{j=1}^{J}\underset{\text{AP-RIS-user}}{\underbrace{\mathbf{f}_{nk}^{jm}\bm{\Theta}^j\mathbf{G}^{ij}}},
\end{equation}
where $\widetilde{\mathbf{h}}_{nk}^{im} \in \mathbb{C}^{1\times N_{\mathrm{A}}}$, $\mathbf{f}_{nk}^{jm} \in \mathbb{C}^{1\times L}$, and $\mathbf{G}^{ij} \in \mathbb{C}^{L\times N_{\mathrm{A}}}$ denote the direct spatial channels of AP $i$-user $U_{nk}^{m}$, RIS $j$-user $U_{nk}^{m}$, and AP $i$-RIS $j$ links, respectively.

We assume LoS paths always exist between the APs and its neighboring RISs.
However, the LoS paths for the AP-user and RIS-user links may be blocked.
Therefore, channel $\mathbf{\widetilde{h}}_{nk}^{im}$ can be modeled by
\begin{equation}
\begin{split}
& \mathbf{\widetilde{h}}_{nk}^{im} \!=\! \mathbbm{1}_{nk}^{im,\mathrm{LoS}}\sqrt{N_{\mathrm{A}}}\sqrt{L_{\mathrm{path}}(f,d_{nk}^{im,\mathrm{AU}})}\varUpsilon\bm{\alpha}^{H}\left(\varphi_{nk}^{im0}\right) \\
    & + \sum_{l=1}^{n_{\mathrm{NL}}}\sqrt{N_{\mathrm{A}}}\sqrt{L_{\mathrm{path}}(f,d_{nk}^{im,\mathrm{AU}})}\lambda_{nk}^{iml}(f)\varUpsilon\bm{\alpha}^{H}\left(\varphi_{nk}^{iml}\right),
\end{split}
\end{equation}
where $\varUpsilon$ is the antenna gain, $L_{\mathrm{path}}(f,d_{nk}^{im,\mathrm{AU}})$ is the path loss determined by the frequency $f$ and the distance $d_{nk}^{im,\mathrm{AU}}$ between AP $i$ and user $U_{nk}^{m}$.
$\mathbbm{1}_{nk}^{im,\mathrm{LoS}}\in\{0,1\}$ indicates the existence of the LoS path, which is determined by the LoS probability based on the indoor THz blockage model \cite{Blocakge_Wu_2021}.
Moreover, $n_{\mathrm{NL}}$ signifies the number of NLoS paths, and $\lambda_{nk}^{iml}$ is the reflection coefficient for NLoS path $l$ \cite{NLoSReflection_Piesiewicz_2007}.
$\varphi_{nk}^{iml}$ denotes the angle of departure (AoD) for the downlink channel between AP $i$ and user $U_{nk}^{m}$. Given AoD $\varphi$, the array response vector can be denoted by $\bm{\alpha}(\varphi) = \frac{1}{\sqrt{N_{\mathrm{A}}}}\left[1,...,e^{j\pi[m\sin\varphi]},..., e^{j\pi[(N_{\mathrm{A}}-1)\sin\varphi]}\right]^T$.
According to  \cite{LoSGain_Jornet_2011}, the path loss $L_{\mathrm{path}}(f,d)$ includes both spreading loss and absorption loss, i.e.,
\begin{equation}
\begin{split}
L_{\mathrm{path}}(f,d)[dB]
&\!=\!L_{\mathrm{spread}}(f,d)[dB]\!+\!L_{\mathrm{absorption}}(f,d)[dB] \\
&\!=\!20\log_{10}\!\left(\frac{c}{4\pi fd}\right)\!\!-10k(f)d\log_{10}e,
\end{split}
\end{equation}
where $k(f)$ is the frequency-dependent medium absorption coefficient, $d$ is the distance, and $c$ is the light speed.

\subsection{User Clustering}
In traditional CSI-based massive MIMO-NOMA networks, the user clustering usually contains a cluster-head selection (CHS) procedure that chooses cluster head based on channel conditions, followed by which the remaining users are grouped with the highest channel-correlated cluster heads \cite{CSI_CHS_2019}.
However, the CSI-based user clustering is inflexible to guarantee the ultra-high data rate and mitigate interference for SE users.
To achieve customizable and intelligent communication experiences, we extend the traditional CSI-based user clustering to a QoS-based user clustering.
Relying on the smartly reconfigured beams, users' spatial channels can be flexibly aligned to achieve adaptive user clustering.
Specifically, we group multiple IoT users with a single SE user. In this way, we can ensure multiplexing gain for SE users while enhancing IoT user connections.
In light of this, we assume the number of RF chains activated at each AP is equal to its serving SE users, i.e., $N_{\mathrm{R}}=K_{\mathrm{S}}$.
Let $U_{n1}^{m}$ represent the SE user, and $U_{nk}^{m}, \forall k>1$, be the IoT user.
The low-complexity user clustering scheme based on the reconfigured channels can be stated as follows.

The SE users are firstly selected as the cluster heads of different clusters.
Then, we define the channel spatial correlation $\mathcal{C}\left(\mathbf{h}_{k}^{mm},\mathbf{h}_{n1}^{mm}\right)$ between an ungrouped IoT user $k$ and the cluster head $U_{n1}^{m}$ as
\begin{equation}\label{channel_corr}
\mathcal{C}\left(\mathbf{h}_{k}^{m},\mathbf{h}_{n1}^{m}\right) = \frac{\left|\mathbf{h}_{n1}^{m}\left(\mathbf{h}_{k}^{m}\right)^{H}\right|}{\|\mathbf{h}_{k}^{m}\|\|\mathbf{h}_{n1}^{m}\|}.
\end{equation}
Thereafter, multiple IoT users can be non-orthogonally grouped into the same cluster, where the cluster head achieves the strongest reconfigured channel correlations with them.
The computational complexity of the  QoS-based clustering scheme is $\mathcal{O}\left(K_{\mathrm{U}}N_{\mathrm{R}}\right)$. In comparison, the conventional CSI-based clustering scheme has the computational complexity of $\mathcal{O}\left(N_{\mathrm{R}}\left(K_{\mathrm{U}}+K_{\mathrm{S}}\right)^{2}\right)$ \cite{CSI_CHS_2019}.



\subsection{Customized NOMA Decoding}
Relying on the adaptively aligned channels through RISs, the NOMA user decoding order in each cluster can be flexibly customized based on QoS requirements without degrading the system performance.
Exploiting the NOMA successive interference cancellation (SIC), SE users can subtract signals of other users within the same cluster, and completely eliminate intra-cluster interference based on power-domain multiplexing to realize ultra-high-speed and reliability-guaranteed communications.
To realize this goal, each SE user with the highest QoS requirements in its cluster would be decoded at last, while the data signals of $K_n^m-1$ IoT users in each cluster $n$ would be decoded first.

To ensure the SIC at each IoT users in the same cluster can be successfully carried out, the NOMA decoding order of IoT users are rearranged according to equivalent channel gains as $|\mathbf{h}_{n2}^{m}\mathbf{V}^m\mathbf{w}_{n}^{m}|^2 \geq |\mathbf{h}_{n3}^{m}\mathbf{V}^m\mathbf{w}_{n}^{m}|^2 \geq ... \geq |\mathbf{h}_{nK_n^m}^{m}\mathbf{V}^m\mathbf{w}_{n}^{m}|^2$.
Thereafter, the SE user would sequentially decode the signals of the former $K_n^m-1$ IoT users to completely cancel intra-cluster interference.
To successfully cancel interference from IoT user $U_{nk}^{m}, \forall k>1$, imposed to SE user $U_{n1}^{m}$, we have the following SIC constraint
\begin{equation}\label{SIC_constraint}
\gamma_{n1 \to nk}^{m} \geq \gamma_{nk \to nk}^{m}, \forall k > 1,
\end{equation}
where  $\gamma_{n,k \to n,k}^{m}$ is the signal-to-interference-and-noise ratio (SINR) of IoT user $U_{nk}^{m}$ for decoding its signal, and $\gamma_{n,1 \to n,k}^{m}$  is the SINR to decode the signal of IoT user $U_{nk}^{m}$ at SE user $U_{n1}^{m}$.
We denote $\mathbbm{1}_{nk}^{m,\mathrm{fail}} = 0$ if the SIC constraint \eqref{SIC_constraint} can be satisfied, i.e., the SE user $U_{n1}^{m}$ can cancel the intra-cluster interference from IoT user $U_{nk}^{m}$.
Otherwise $\mathbbm{1}_{nk}^{m,\mathrm{fail}} = 1$, and the intra-cluster interference from IoT user $U_{nk}^{m}$ would be taken as noise at the SE user.
In this way, $\gamma_{n,1 \to n,k}^{m}$ can be expressed as
\begin{equation}\label{SINR_decode}\small
\begin{split}
&\gamma_{n,1 \to n,k}^{m} \!=\!
{\left|\mathbf{h}_{n1}^{mm}\mathbf{V}^{m}\mathbf{w}_{n}^{m} \right|^{2} \! p_{nk}^{m}} \! \times \!
\! \bigg[ \! \bigg(\sum_{k'=1}^{k-1}p_{nk'}^{m} \!\!+\!\! \sum_{k'>k} \! \mathbbm{1}_{nk'}^{m,\mathrm{fail}}p_{nk'}^{m}\bigg) \! \\&
\left|\mathbf{h}_{n1}^{mm}\mathbf{V}^{m}\mathbf{w}_{n}^{m}\right|^{2}
+ \sum_{(m',n')\atop\ne(m,n)}\left|\mathbf{h}_{n1}^{m'm}\mathbf{V}^{m'}\mathbf{w}_{n'}^{m'}\right|^{2}P_{n'}^{m'}
+\sigma^{2}\bigg]^{-1},
\end{split}
\end{equation}
where $\sigma^{2}$ is the power spectral density of additional white Gaussian noise (AWGN),  $p_{nk}^{m}$ is the power allocation coefficient for user $U_{nk}^{m}$, and $P_{n}^{m} = \sum_{k=0}^{K_n^{m}} p_{nk}^{m}$ is the power consumption of RF chain $n$ at AP $m$.

%

\subsection{Sub-Connected Hybrid Beamforming}
We design the analog beamforming at each AP to increase effective data gain of each cluster, and design the digital beamforming for suppressing residual inter-cluster interference that is not mitigated by RISs based on zero-forcing (ZF).

The analog beamforming matrix with the sub-connected structure can be formulated as
\begin{equation}
\mathbf{V}^m\!=\!\left[\!\begin{array}{ccc}
\mathbf{v}_{m11}^{\mathrm{sub}} & ... & \mathbf{v}_{m1N}^{\mathrm{sub}}\\
\mathbf{v}_{m21}^{\mathrm{sub}} & ... & \mathbf{v}_{m2N}^{\mathrm{sub}}\\
... & ... & ...\\
\mathbf{v}_{mN1}^{\mathrm{sub}} & ... & \mathbf{v}_{mNN}^{\mathrm{sub}}
\end{array}\!\right]\!=\!\left[\!\begin{array}{cccc}
\mathbf{v}_{m11}^{\mathrm{sub}} &  ... & \bm{0}\\
\bm{0} &  ... & \bm{0}\\
... & ... & ...\\
\bm{0} & ... & \mathbf{v}_{mNN}^{\mathrm{sub}}
\end{array}\!\right]\!,
\end{equation}
where $\mathbf{v}_{mni}^{\mathrm{sub}} = \mathbf{0}$ for $n \ne i$, and $\mathbf{v}_{mnn}^{\mathrm{sub}} \in \mathbb{C}^{N_{\mathrm{sub}}\times 1}$ has an amplitude $\frac{1}{\sqrt{N_{\mathrm{sub}}}}$.
The sub-connected analog beamforming is designed to improve the effective gains toward cluster heads that are highly spatial-correlated to the cluster members.
Considering $B^{\mathrm{A}}$-bits quantized phase shifters in each antenna subarray, the $i$-th element of the active analog beamforming vector $\mathbf{v}_{mnn}^{\mathrm{sub}}$ is given by \cite{THzMIMONOMA_Zhang_2020,AdaptiveHP_Zhu_2016}
\begin{equation}
\mathbf{v}_{mnn}^{\mathrm{sub}}(i) = \frac{1}{\sqrt{N_{\mathrm{sub}}}} e^{-j \frac{2\pi \varpi_{mn}^{*}}{2^{B^{\mathrm{A}}}}}, i \in \{1,2,..,N_{\mathrm{sub}}\},
\end{equation}
where $\varpi_{mn}^{*}$ is the quantized phase calculated by
\begin{equation}\label{quantized_phase}
\varpi_{mn}^{*} = {\arg \min}_{\varpi \in \{0,1,...,2^{B^{\mathrm{A}}}-1\}} \left|e^{j\frac{2\pi \varpi}{2^{B^{\mathrm{A}}}}} - \frac{\mathbf{h}_{n1}^{mm}(i)}{\left|\mathbf{h}_{n1}^{mm}(i)\right|} \right|.
\end{equation}

To further suppress the effective inter-cluster interference that hasn't been canceled by RISs, we design the ZF digital beamforming as follows.
Let $\mathbf{w}_n^m  \in \mathbb{C}^{N_{\mathrm{R}}\times 1}$ denote the digital beamforming vector for cluster $n$ at AP $m$.
The cluster center is given as
$\widehat{\mathbf{h}}_n^{mm}= \frac{1}{K_n^m}\sum_{k \in \mathcal{K}_n^m}\left(\mathbf{h}_{nk}^{mm}\mathbf{V}^{m}\right)^{T} $.
Denote $\widehat{\mathbf{H}}^{mm}  = \left[\widehat{\mathbf{h}}_1^{mm},\widehat{\mathbf{h}}_2^{mm}, ...,\widehat{\mathbf{h}}_{N_{\mathrm{R}}}^{mm}\right]$.
Therefore, the ZF digital beamforming is calculated as
\begin{equation}
\begin{split}
\widehat{\mathbf{W}}^{m}
&= \left[\widehat{\mathbf{w}}_{1}^{m}, ..., \widehat{\mathbf{w}}_{N_{\mathrm{R}}}^{m}\right] \\
&= \left(\widehat{\mathbf{H}}^{mm}\right)^{H}\left[\widehat{\mathbf{H}}^{mm}\left(\widehat{\mathbf{H}}^{mm}\right)^{H}\right]^{-1}.
\end{split}
\end{equation}

By introducing column power normalizing, the baseband precoding matrix can be expressed as
\begin{equation}
\mathbf{w}_{n}^{m} = \frac{\widehat{\mathbf{w}}_{n}^{m}}{\|\mathbf{V}^{m}\widehat{\mathbf{w}}_{n}^{m}\|_2}.
\end{equation}

\subsection{Transmission Rate}
Define $\mathbb{E}\{{s}_{nk}^{m}\left({s}_{nk}^{m}\right)^{H}\} = 1$ as the transmitted symbols with normalized power, and $x_{n}^{m} = \sum_{k=0}^{K_n^m}\sqrt{p_{nk}^{m}}s_{nk}^{m}$ as the transmitted signal in cluster $n$.
Define $z_{nk}^{m}$ as the AWGN.
The baseband signal received by each SE user $U_{n1}^{m}$ after SIC can be formulated as
\begin{equation}\small
\begin{split}
y_{n1}^{m}
&\! = \! \underset{\text{desired signal}}{\underbrace{\mathbf{h}_{n1}^{mm}\mathbf{V}^m\mathbf{w}_{n}^{m} \sqrt{p_{n1}^{m}}s_{n1}^{m}}}
+ \! \underset{\text{residual intra-cluster interference}}{\underbrace{\mathbf{h}_{n1}^{mm}\mathbf{V}^m\mathbf{w}_{n'}^{m} \! \! \sum_{k>1} \! \mathbbm{1}_{nk}^{m,\mathrm{fail}} \! \sqrt{p_{nk}^{m}}s_{nk}^{m}}}\\&
+\! \underset{\text{inter-cluster interference}}{\underbrace{\sum_{(m',n') \ne (m,n)}\mathbf{h}_{n0}^{m'm}\mathbf{V}^{m'}\mathbf{w}_{n'}^{m'}x_{n'}^{m'}}}
+\underset{\text{noise}}{\underbrace{z_{n1}^{m}}}.
\end{split}
\end{equation}

Therefore, the SINR of each SE user $U_{n1}^{m}$ can be written as
\begin{equation}\label{SINR_UHQ}
\begin{split}
\gamma_{n1}^{m}&=
{\left|\mathbf{h}_{n1}^{mm}\mathbf{V}^{m}\mathbf{w}_{n}^{m}\right|^{2}p_{n1}^{m}}
\bigg(\sum_{k>1}\mathbbm{1}_{nk}^{m,\mathrm{fail}}\left|\mathbf{h}_{n1}^{mm}\mathbf{V}^{m}\mathbf{w}_{n}^{m}\right|^{2}p_{nk}^{m}\\&
+\sum_{(m',n')\ne(m,n)}\left|\mathbf{h}_{n1}^{m'm}\mathbf{V}^{m'}\mathbf{w}_{n'}^{m'}\right|^{2}P_{n'}^{m'}+\sigma^{2}\bigg)^{-1}.
\end{split}
\end{equation}

Moreover, the SINR of IoT users $U_{nk}^{m}, \forall k>1$, can be given by
\begin{equation}\label{SINR_IoT}
\begin{split}
\gamma_{nk}^{m}&=
\left|\mathbf{h}_{nk}^{mm}\mathbf{V}^{m}\mathbf{w}_{n}^{m}\right|^{2}p_{nk}^{m}
\bigg(
\sum_{k'=1}^{k-1}\left|\mathbf{h}_{nk}^{mm}\mathbf{V}^{m}\mathbf{w}_{n}^{m}\right|^{2}p_{nk'}^{m}+\\&
\sum_{(m',n')\ne(m,n)}\left|\mathbf{h}_{nk}^{m'm}\mathbf{V}^{m'}\mathbf{w}_{n'}^{m'}\right|^{2}p_{n'}^{m'}+\sigma^{2}
\bigg)^{-1}.
\end{split}
\end{equation}

Hence, the data rate of each user $U_{nk}^{m}$ at each time slot can be expressed as
\begin{equation}\label{rate}
R_{nk}^{m} = \log_2 \left(1+ \gamma_{nk}^{m}\right), \forall k \in \mathcal{K}.
\end{equation}

\subsection{Power Consumption}\label{Sec_Power}
The total power consumption for the proposed networks at each time slot includes both transmit and circuit power, which can be formulated as
\begin{equation}
P \!=\! \sum_{m=1}^{M}\sum_{n=1}^{N_{\mathrm{R}}} \xi P_{n}^{m} + KP_{\mathrm{D}} + M P_{\mathrm{AP}}
+ \sum_{j=1}^{J}\sum_{l=1}^{L} \omega_{l}^{j}P_{\mathrm{RIS}}\left(B^{\mathrm{R}}\right),
\end{equation}
where $\xi$ denotes the inefficiency of the phase shifter in the THz network,
$P_{\mathrm{D}}$ is the circuit power consumption at each user,
and $P_{\mathrm{RIS}}\left(B^{\mathrm{R}}\right)$ denotes hardware energy dissipated at each RIS element depending on the RIS resolution bit $B^{\mathrm{R}}$\cite{EERIS_Huang_2019}.
Moreover, $P_{\mathrm{AP}}$ is the circuit power consumption at each AP, which is given by
\begin{equation}
P_{\mathrm{AP}}=P_{\mathrm{BB}}+N_{\mathrm{R}}P_{\mathrm{RF}}+N_{\mathrm{A}}\left(P_{\mathrm{PS}}+P_{\mathrm{A}}\right),
\end{equation}
where $P_{\mathrm{BB}}$ is the baseband power consumption, and $P_{\mathrm{RF}}, P_{\mathrm{PS}}$, and $ P_{\mathrm{A}}$ denote power consumption of per RF chain, per phase shifters and per power amplifies, respectively.

Therefore, the network energy efficiency $\eta$ at each time slot can be formulated as
\begin{equation}\label{EE}\small
\eta \!=\! \frac{\sum_{m=1}^{M}\sum_{n=1}^{N_{\mathrm{R}}}\sum_{k=1}^{K_n^m}\log_2\left(1+\gamma_{nk}^{m}\right)}
{\sum_{m=1}^{M}\sum_{n=1}^{N} \xi P_{n}^{m} \!+\! KP_{\mathrm{D}} \!+\! MP_{\mathrm{AP}} \!+\! \sum_{j=1}^{J}\sum_{l=1}^{L} \omega_{l}^{j}P_{\mathrm{RIS}}\left(B^{\mathrm{R}}\right)}.
\end{equation}

\subsection{Reliability Model}
Suppose the THz MIMO-NOMA networks maintain a traffic buffer queue $q_{nk}^{m}(t)$ for each user, $\forall t\in\mathcal{T}$, which aims to transmit specific data volume in a predefined time. At the beginning of each time slot $t$, the queue length $q_{nk}^{m}(t)$ of user $U_{nk}^{m}$ can be updated by
\begin{equation}
q_{nk}^{m}(t+1) = A_{nk}^{m}(t) + \left[ q_{nk}^{m}(t) - R_{nk}^{m}(t) \right]^{+},
\end{equation}
where $R_{nk}^{m}(t)$ is the transmission data rate of \eqref{rate},
$A_{nk}^{m}(t)$ denotes the traffic arrival column at time slot $t$ following the poisson distribution,
and $\left[x\right]^{+} = \max\{x,0\}$.

To ensure reliable transmission, we should guarantee the queue stability and keep the outages below a predefined threshold \cite{DFLReliability_Samarakoon_2019}.
The queue stability is ensured by
\begin{equation}\label{queue}
\mathbb{E}\left[q_{nk}^{m}(t)\right] = \lim_{T \to \infty} \sum _{t=1}^{T} q_{nk}^{m}(t) < \infty, \forall k \in \mathcal{K}.
\end{equation}
Moreover, the outage probability, i.e., the probability that the traffic queue length $q_{nk}^{m}(t)$ of device $U_{nk}^{m}$ exceeds the threshold $q_{nk}^{m,\max}$, is limited by a certain threshold $\epsilon_{nk}^{m}$.
Therefore, the reliability conditions can be formulated as
\begin{equation}\label{reliability}
\mathrm{Pr}\left(q_{nk}^{m}(t)\geq q_{nk}^{m,\max}\right)\leq\epsilon_{nk}^{m}, \forall k \in \mathcal{K}.
\end{equation}

\section{Lyapunov Optimization Based Dec-POMDP Problem}
In this Section, we first formulate the constrained Dec-POMDP problem, and then transfer the constrained Dec-POMDP problem into a normal Dec-POMDP problem based on the Lyapunov optimization theory.

\subsection{Constrained Dec-POMDP Model}
We aim to dynamically find a stationary online policy $\pi$, which can jointly optimize the dynamic RIS element selection, coordinated discrete phase-shift control, and transmit power allocation by observing the environment state at each time slot $t$.
Under the given user clustering, NOMA decoding, and hybrid beamforming schemes, the joint policy dynamically maximizes the expected system energy efficiency for the smart reconfigurable THz MIMO-NOMA network, while satisfying various QoS requirements in terms of data rate and transmission reliability of users. Therefore, the online objective function can be formulated as follows
\begin{subequations}\label{P1}
\begin{equation*}
\mathcal{P}_{0}: \max_{\mathbf{a}(t)} \lim_{T\to \infty} \sum_{t=0}^{T} \mathbb{E} \left[\eta(t)\right], \tag{\ref{P1}{a}}
\end{equation*}
\begin{align*}
{\mathrm{s.t.}}~ & R_{nk}^{m}(t) \geq R_{nk}^{m,\min}, \forall k \in \mathcal{K}, t \in \mathcal{T}, \label{constraint_rate} \tag{\ref{P1}{b}}
\\&
\sum_{n=1}^{N_{\mathrm{R}}} \sum_{k=1}^{K_{n}^{m}(t)} p_{nk}^{m}(t) \leq P_{\mathrm{AP}}^{\max}, \forall m \in \mathcal{M}, t \in \mathcal{T}, \label{constraint_power} \tag{\ref{P1}c}
\\&
\omega_{l}^{j}(t) \in\{0, 1\},\forall j \in \mathcal{J}, 1 \leq l \leq L, t \in \mathcal{T},  \label{constraint_bin} \tag{\ref{P1}d}
\\&
\beta_{l}^{j}(t) \in \{0,1,...,2^{B^{\mathrm{R}}-1}\}, \forall j \in \mathcal{J}, 1 \leq l \leq L, t \in \mathcal{T},  \label{constraint_RIS}  \tag{\ref{P1}e}
\\&
\eqref{queue}, \eqref{reliability}, \tag{\ref{P1}f}
\end{align*}
\end{subequations}
where $\mathbf{a}(t) = \left[\bm{\omega}(t), \bm{\alpha}(t), \bm{\beta}(t)\right]$.
Constraint \eqref{constraint_rate} ensures the minimum data rate of each user,
\eqref{constraint_power} denotes the total transmit power should be less than the maximum power $P_{\mathrm{AP}}^{\max}$,
\eqref{constraint_bin} signifies the binary ON/OFF state of RIS elements,
and constraint \eqref{constraint_RIS} represents the discrete phase-shift control.
Note that the SIC constraint \eqref{SIC_constraint} is ignored since it has been implied in the SINR expression \eqref{SINR_UHQ} via the indicator $\mathbbm{1}_{nk}^{m,\mathrm{fail}}$.

This online policy can be modeled as a constrained Markov Decision Process (MDP).
However, solving $\mathcal{P}_{0}$ in a centralized way is computationally inefficient due to the large-scale joint state-action space and the heavy overhead of high-dimensional information exchange from  multiple APs and RISs to the centralized controller.
To tackle $\mathcal{P}_{0}$ in an efficient and low-complexity manner, we model the long-term energy efficiency optimization problem as a constrained Dec-POMDP.
In detail, the POMDP provides a generalized framework to describe an MDP with incomplete information, while Dec-POMDP extends POMDP to the decentralized settings as $\mathcal{P}=\left(\mathcal{I},\mathcal{S}, \Omega,\mathcal{A},P_{\mathbf{s}\to\mathbf{s}'}^{\mathbf{a}},\pi,r,\Gamma\right)$.
Here, $\mathcal{I}$ denotes a set of agents. $\mathcal{S}$, $\Omega$, and $\mathcal{A}$ denote the state, observation, and action spaces.
At each slot $t$, given the true state $\mathbf{s}(t)\in\mathcal{S}$, each agent $i$ partially observes a local observation, based on which it can determine a local action using  decentralized policy $\pi$.
The joint observation and action vectors of all agents are denoted by $\mathbf{\Omega}(t)\in \Omega$ and $\mathbf{a}(t)\in\mathcal{A}$, and the decentralized policy $\pi(\mathbf{a}(t)|\mathbf{s}(t))$ means the probability of taking joint action $\mathbf{a}(t)$ at the joint state $\mathbf{s}(t)$.
Furthermore, $P_{\mathbf{s}\to\mathbf{s}'}^{\mathbf{a}} = \mathbb{P}(\mathbf{s}'|\mathbf{s},\mathbf{a})$ is the probability of transferring into a new state $\mathbf{s}'$ by taking action $\mathbf{a}$ at state $\mathbf{s}$.
Given the global reward function $r\left(\mathbf{s}(t),\mathbf{a}(t)\right)$ and the discount factor $\Gamma$, the agents can cooperate and coordinate to maximize the discounted return.
In this work, we model APs and RISs as two types of agents, whose set is denoted as $\mathcal{I} = \mathcal{I}^{0} \cup \mathcal{I}^{1}$.
Let $\mathcal{I}^{u}$ denote the set of the $u$-th type agents, where $u = \{0,1\}$ represent APs and RISs respectively.
We specify the formulated Dec-PoMDP as follows.

\subsubsection{Observation}
The joint observation can be expressed as $\mathbf{\Omega}(t) = \left[\mathbf{\Omega}_{1}^{0}(t),...,\mathbf{\Omega}_{M}^{0}(t),\mathbf{\Omega}_{1}^{1}(t),...,\mathbf{\Omega}_{J}^{1}(t)\right]$, where
$\mathbf{\Omega}_{i}^{u}(t)$ is the local observation of agent $i \in \mathcal{I}^{u}$ that involves part of the environment information, given by
\begin{equation}\label{state_P1_AP}\small
      \mathbf{\Omega}_{i}^{0}(t) \!=\! \bigg[\! \left\{\bm{\widetilde{\mathrm{H}}}^{ii'}(t)\right\}_{i'\in\mathcal{N}_{i}^{0}\cup\{i\}},
            \mathbf{q}^{i}(t), \bm{\tau}_{i}^{0}(t) \! \bigg],
            \forall i \in \mathcal{I}^{0},
\end{equation}
\begin{equation}\label{local_obs_RIS}\small
        \mathbf{\Omega}_{i}^{1}(t) = \bigg[\left\{\mathbf{F}^{im}(t)\right\} _{m\in\mathcal{N}_{i}^{1}},
        \left\{ \mathbf{G}^{mi}(t)\right\} _{m\in\mathcal{N}_{i}^{1}}\!,
            \bm{\tau}_{i}^{1}(t) \bigg],
            \forall i \in \mathcal{I}^{1},
\end{equation}
where $\bm{\tau}_{i}^{u}(t)$ denotes the local action-observation history of agent $i$, and $\mathcal{N}_{i}^{0}$ and $\mathcal{N}_{i}^{1}$ denotes the neighboring agent sets of RIS $i\in \mathcal{I}^{0}$ and AP $i\in \mathcal{I}^{1}$, respectively.
Furthermore, we define $\widetilde{\mathbf{H}}^{ii'}(t)=\left[\left\{\widetilde{\mathbf{h}}_{nk}^{ii'}(t)\right\}_{k\in\mathcal{K}^{i'}}\right]$ and $\mathbf{F}^{im}(t)=\left[\left\{\mathbf{f}_{nk}^{im}(t)\right\}_{k\in\mathcal{K}^{m}}\right]$ as the vectorized channels.
Note that in our settings, the joint observation also represents the true environment state, i.e., $\mathbf{\Omega}(t) = \mathbf{s}(t)$.

\subsubsection{Action}
In this work, we consider a long-term joint RIS element selection, coordinated discrete phase-shift control, and power allocation optimization problem. Therefore, we define the joint action vector as $\mathbf{a}(t) = [\mathbf{a}_{1}^{0}(t),..., \mathbf{a}_{M}^{0}(t), \mathbf{a}_{1}^{1}(t), ..., \mathbf{a}_{J}^{1}(t)]$,
where $\mathbf{a}_{i}^{u}(t)$ is the local action of agent $i\in\mathcal{I}^{u}$ that can be expressed as
\begin{equation}
\mathbf{a}_{i}^{u}(t)=\begin{cases}
\mathbf{p}^{i}(t), & \forall i\in\mathcal{I}^{0},\\
\left[\bm{\omega}^{i}(t),\bm{\beta}^{i}(t)\right], & \forall i\in\mathcal{I}^{1}.
\end{cases}
\end{equation}
Here, $\mathbf{p}^{i}(t) = \left[p_{11}^{m}(t), p_{12}^{m}(t),...,p_{N_{\mathrm{R}}K_{N_{\mathrm{R}}}^{m}(t)}^{m}(t)\right]$ is the downlink power allocation vector of AP agent $i\in\mathcal{I}^{0}$,
and $\bm{\omega}^{i}(t) = \left[\omega_{1}^{i}(t),\omega_{2}^{i}(t),...,\omega_{L}^{i}(t)\right]$ and $\bm{\beta}^{i}(t)=\left[\beta_{1}^{i}(t),\beta_{2}^{i}(t),...,\beta_{L}^{i}(t)\right]$ denote the dynamic reflecting element selection and the discrete phase-shift control of RIS agent $i\in\mathcal{I}^{1}$, respectively.

\subsubsection{Decentralized Policy}
The joint policy is decomposed into multiple decentralized and parallel local policies $\pi(\mathbf{a}_{i}^{u}(t)|\mathbf{\Omega}_{i}^{u}(t))$ for each agent $i\in\mathcal{I}^{u}$.
At each time slot $t$, all the agents take local actions $\mathbf{a}_{i}^{u}(t)$ based on the observed local observations $\mathbf{\Omega}_{i}^{u}(t)$ to cooperatively maximize the global reward.

\subsubsection{Global Reward}
Let $\delta^{m}(t)\! \!=\! \sum_{n=1}^{N_{\mathrm{R}}}\sum_{k=1}^{K_n^m(t)}\! \left[R_{nk}^{m,\min}(t)\!-\!R_{nk}^{m}(t)\right]^{+}$ be the data rate constraint violation.
Then, the global reward can be defined as $r\left(\mathbf{s}(t),\mathbf{a}(t)\right)=\eta(t) - \xi\delta(t)$, which maximizes the network energy efficiency as well as satisfying data rate requirements. Here, $\delta(t) =  \sum_{m=1}^{M}\delta^{m}(t)$, and $\xi$ is a non-negative parameter that imposes a penalty for data rate violation.
However, due to the intractable long-term average queue stability and reliability constraints, it is infeasible to directly use the general MADRL methods to solve the constrained Dec-POMDP problem \cite{CMDP_Khairy_2020,CMDPLya_Wu_2020}.
Hence, we recast the constrained Dec-POMDP into a standard Dec-POMDP based on the Lyapunov optimization theory \cite{Lyapunov_Neely_2010}.

\subsection{Equivalently Transferred Dec-POMDP}
Based on the Markov's inequality \cite{Probability_Ghosh_2002}, we can obtain the upper bound of the outage probability $\mathrm{Pr}\left(q_{nk}^{m}(t)\geq q_{nk}^{m,\max}\right)\leq \mathbb{E}\left[q_{nk}^{m}(t)\right]/q_{nk}^{m,\max}$.
Therefore, the reliability constraint \eqref{reliability} can be strictly guaranteed by ensuring
\begin{equation}\label{reliability_markov}
\mathbb{E}\left[q_{nk}^{m}(t)\right] \leq q_{nk}^{m,\max} \epsilon_{nk}^{m}, \forall k \in \mathcal{K}.
\end{equation}

To tackle the long-term average constraints \eqref{queue} and \eqref{reliability_markov},
we apply the virtual queue technique and Lyapunov optimization theory \cite{Lyapunov_Neely_2010} to recast the intractable constrained Dec-POMDP as a normal Dec-POMDP, which can be solved by general MADRL algorithms.
To meet the reliability constraints \eqref{reliability_markov}, we construct a virtual queue $Y_{nk}^{m}(t)$ to model the instantaneous deviation of queue length constraints for each user $U_{nk}^{m}$.
Here, $Y_{nk}^{m}$ evolves as follows
\begin{equation}\label{virtual_queue}
Y_{nk}^{m}(t+1) = \left[Y_{nk}^{m}(t) + \left(q_{nk}^{m}(t+1) - q_{nk}^{m,\max} \epsilon_{nk}^{m} \right) \right]^{+},
\end{equation}
where we initialize $Y_{nk}^{m}(t) = 0$ at $t=1$, $\forall k \in \mathcal{K}$.

Let $\bm{\Xi}_{nk}^{m}(t)=\left[q_{nk}^{m}(t), Y_{nk}^{m}(t)\right]$, $\forall k \in \mathcal{K}$. Define $\bm{\Xi}(t) = \left[\bm{\Xi}_{11}^{1}(t), \bm{\Xi}_{12}^{1}(t), ...,\bm{\Xi}_{N_{\mathrm{R}}K^{M}}^{M}(t)\right]$ as the combined queue vector.
Then, we introduce the Lyapunov function $\mathcal{L}(\bm{\Xi}(t)) = \frac{1}{2} \bm{\Xi}(t)\bm{\Xi}^{T}(t)$ to measure the satisfaction status of the reliability constraint, which should be maintained under a low value.
Moreover, the one-step Lyapunov drift is defined as $\Delta\mathcal{L} (t) = \mathcal{L}(\bm{\Xi}(t+1)) -  \mathcal{L}(\bm{\Xi}(t))$, which can be upper bounded by the following Lemma.
\begin{lemma}\label{Lemma_LyapunovDrift}
Define $\Delta{\mathcal{L}}_{nk}^{m}(t) = \mathcal{L}(\bm{\Xi}_{nk}^{m}(t+1)) -  \mathcal{L}(\bm{\Xi}_{nk}^{m}(t))$, then $\Delta{\mathcal{L}}_{nk}^{m}(t)$ can be upper bounded by
\begin{equation}\label{LyapunovDrift}
   \Delta{\mathcal{L}}_{nk}^{m}(t) \! \leq\! C_{nk}^{m} +B_{nk}^{m}(t) + \! \Lambda_{nk}^{m}(t) \left(A_{nk}^{m}(t)-R_{nk}^{m}(t)\right),
\end{equation}
where
$C_{nk}^{m} = 2C_{nk}^{m,\mathrm{q}}+C_{nk}^{m,\mathrm{Y}}$ is a constant, with $C_{nk}^{m,\mathrm{q}}=\frac{1}{2}\left(A_{nk}^{m,\max}\right)^{2}+\frac{1}{2}\left(R_{nk}^{m,\max}\right)^{2}$ and
$C_{nk}^{m,\mathrm{Y}}=\frac{1}{2}\left(q_{nk}^{m,\max}\epsilon_{nk}^{m}\right)^{2}$.
Moreover, $B_{nk}^{m}(t)=\frac{1}{2}\left(q_{nk}^{m}(t)\right)^{2}+Y_{nk}^{m}(t)\left[A_{nk}^{m}(t)+q_{nk}^{m}(t)\right]$ and $\Lambda_{nk}^{m}(t)= Y_{nk}^{m}(t)+2q_{nk}^{m}(t)$ are fixed values at each time slot.
\begin{proof}
See Appendix \ref{Proof_LyapunovDrift}.
\end{proof}
\end{lemma}
Based on the Lyapunov optimization, the original constrained Dec-POMDP of maximizing the system energy efficiency while ensuring the long-term reliability constraints can be transformed into minimising the following drift-minus-bonus function
\begin{equation}
\begin{split}
& \Delta\mathcal{L}(t) - \zeta \eta(t) + \xi \delta(t) \leq  \sum_{m=1}^{M}\sum_{n=1}^{N_{\mathrm{R}}}\sum_{k=1}^{K_n^m(t)} \bigg[ C_{nk}^{m} + B_{nk}^{m}(t) \\
&  +  \Lambda_{nk}^{m}(t)\left(A_{nk}^{m}(t)-R_{nk}^{m}(t)\right) \bigg]  - \zeta \eta(t) + \xi \delta(t) ,
\end{split}
\end{equation}
where $\zeta$ denotes the positive coefficient that controls the tradeoff between energy efficiency and transmission reliability, and the inequality is implied by Lemma \ref{Lemma_LyapunovDrift} and $\Delta \mathcal{L}(t) = \sum_{m=1}^{M}\sum_{n=1}^{N_{\mathrm{R}}}\sum_{k=1}^{K_n^m(t)} \Delta\mathcal{L}_{nk}^{m}(t)$.
By ignoring the independent terms from control variables, $\mathcal{P}_1$ can be rearranged as
\begin{subequations}\label{P2}\small
\begin{equation}
\max_{\pi} \! \!\lim_{T \to \infty} \frac{1}{T}
 \sum_{t=1}^{T} \mathbb{E}_{\pi} \! \left[ \zeta \eta(t) \!-\! \xi \delta(t) \!+\! \sum_{m=1}^{M}\sum_{n=1}^{N_{\mathrm{R}}}\sum_{k=1}^{K_n^m(t)}\! \Lambda_{nk}^{m}(t)R_{nk}^{m}(t) \right], \tag{\ref{P2}{a}}
\end{equation}
\begin{align*}
{\mathrm{s.t.}}~ & \eqref{constraint_power}-\eqref{constraint_RIS}. \tag{\ref{P2}{b}}
\end{align*}
\end{subequations}

Based on \eqref{P2}, the local observation of AP $i \in \mathcal{I}^{0}$ in \eqref{state_P1_AP} can be transformed as
\begin{equation}\label{local_obs_AP}\small
\mathbf{\Omega}_{i}^{0}(t)\!=\! \bigg[ \left\{\bm{\widetilde{\mathrm{H}}}^{ii'}(t)\right\} _{i'\in\mathcal{N}_{i}^{0}\cup\{i\}}\!,
\bm{\Lambda}^{i}(t), \bm{\tau}_i^{0}(t) \bigg]\!,
\forall i \in \mathcal{I}^{0},
\end{equation}
where $\bm{\Lambda}^{i}(t) = \left[\Lambda_{11}^{i}(t), \Lambda_{12}^{i}(t),...,\Lambda_{N_{\mathrm{R}}K_{n}^{i}(t)}^{i}(t)\right]$.
From \eqref{P2}, the global reward function $r\left(\mathbf{s}(t),\mathbf{a}(t)\right)$ in the equivalent Dec-POMDP is modified into
\begin{equation}\small
r\left(\mathbf{s}(t), \mathbf{a}(t)\right)= \zeta \eta(t) - \xi \delta(t) +
\sum_{m=1}^{M}\sum_{n=1}^{N_{\mathrm{R}}}\sum_{k=1}^{K_n^m(t)} \Lambda_{nk}^{m}(t)R_{nk}^{m}(t).
\end{equation}

\section{\textit{GE-VDAC} Algorithm for Distributed RIS Configuration and Power Allocation}

\subsection{Review of MADRL Methods}
The standard Dec-POMDP problem can be solved based on general cooperative MADRL algorithms \cite{MADRLReview_Nguyen_2020},\cite{MAAC_Ryan_2017}.
By exploiting the remarkable representation ability of deep neural network (DNN) for approximating arbitrary non-convex functions,
MADRL algorithms can learn the mapping from agents' observations to actions through exploitation and exploration.
However, MADRL algorithms commonly suffer partial observability, environment instability, and credit assignment problems \cite{MADRLReview_Nguyen_2020}.
To cope with environment instability, the multi-actor critic framework based on centralized-training and distributed-execution (CTDE) design is proposed \cite{MAAC_Ryan_2017}, which learns distributed policies with a centralized critics.
However, since it utilizes a unified critic value to compute local policy gradients, agents cannot fairly learn to contribute for global optimization, leading to the so-called \textit{multiagent credit assignment} problem \cite{MADRLReview_Nguyen_2020}, \cite{RewardShaping_Wolpert_2002}, also referred to as \textit{indolent agent}.
Therefore, the idea of \emph{different rewards} has been introduced, which computes the individual reward for  each agent to evaluate their contributions and motivate coordinations.
QMIX \cite{QMIX_Rashid_2018} is the widely utilized mechanism that decomposes the joint state-action value into monotonic individual state-action functions for different agents. To overcome the monotonicity limitation of QMIX, QTRAN with higher generalization of value factorization can be leveraged \cite{QTRAN}. Nevertheless, both QMIX and QTRAN as extensions of the deep Q-learning are mainly used to deal with the problems with discrete action spaces.
In \cite{VDAC_Su_2021}, value-decomposition actor-critic (\emph{VDAC}) is introduced to incorporate the value decomposition into the advantage actor-critic (A2C) framework, which has continuous action space and higher sample efficiency.

In our proposed framework, it's critical to facilitate coordination among multiple APs and RISs to mitigate interference, improve system energy efficiency, and guarantee diversified QoS requirements.
Under the realistic partially observable environment, although the joint design of RIS configuration and power allocation has been decomposed into distributed local policies,
information still need to be exchanged among neighboring agents to achieve fully coordination.
However, considering the high-dimensional CSI information resulting from the massive MIMO structure, directly exchanging information among interacting neighboring agents during each execution can cause high communication overhead and latency.
Therefore, the existing MADRL algorithms are still inefficient to solve the highly coupled Dec-POMDP problem.
In this section, we propose \textit{GE-VDAC}, a novel MADRL algorithm that can tackle the \textit{multiagent credit assignment} as well as  the above problems.

\subsection{The Proposed GE-VDAC Framework}
\textit{GE-VDAC} extends the commonly-used CTDE design in the existing MADRL algorithms, which realizes more efficient cooperative learning by integrating two techniques, i.e., graph embedding and \textit{different rewards}.
The interplay among interacting agents are modeled as a directed communication graph.
Instead of directly exchanging high-dimensional information, the neighboring agents exchange low-dimensional embedded features learned by graph embedding, thus requiring fewer information exchange to achieve efficient coordination under partially observable environment.
Moreover, the graph-embedded features possess permutation-invariant property.
By learning distributed DRL policies over the graph-embedded observation space that has reduced dimension and enjoys permutation-invariant property, the learning speed and the generalization ability can be improved.

\begin{figure}
  \centering
  \includegraphics[width=.5\textwidth]{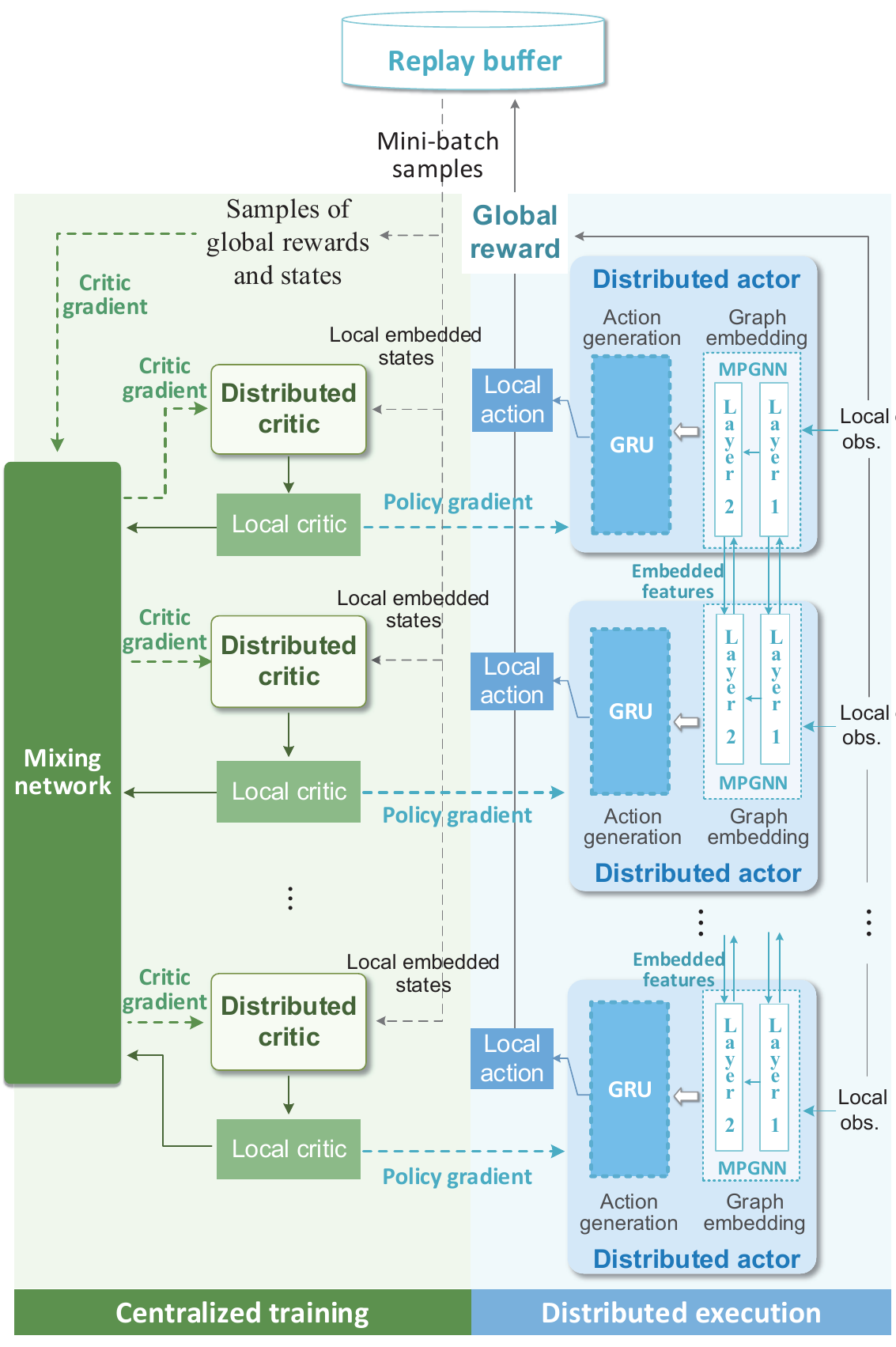}\\
  \caption{Procedure of the proposed \textit{GE-VDAC} algorithm.}\label{fig_alg}
\end{figure}

As shown in Fig. \ref{fig_alg}, the \textit{GE-VDAC} jointly trains three neural networks, i.e., the distributed graph-embedded actors, the distributed critics, and the centralized mixing network.
For simplicity, we assume the agents belonging to the same type $u$ share the neural network parameters of both the distributed actor $\pi^{u}$ and critic $V^{u}$.
The distributed graph-embedded actor contains a semi-distributed \textit{graph embedding module} learning graph-embedded features for efficient information exchange, in conjunction with a fully-distributed \textit{action generation module} that predicts local action based on the local embedded state attained by the \textit{graph embedding module}.
The \textit{graph embedding module} is implemented as message passing graph neural network (MPGNN) \cite{GNN_Xu_2019},\cite{MPGNN_Shen_2021}.
Moreover, the \textit{action generation module} mainly comprises a gated recurrent unit (GRU) \cite{GRU_Cho_2014}, which is a simplified variant of long-short term memory (LSTM) that can achieve comparative performance \cite{GRULSTM_Chung_2014}.
Two fully-connected (FC) layers are connected with the GRU before and after, respectively.
On the other hand, the distributed critic evaluates an individual value of the local embedded state, which achieves differential reward to judge each agent's contribution for global optimization.
Meanwhile, a global mixing network is trained to combine the distributed critics, which guarantees that equivalent policy improvement can be achieved based on the value decomposition.

\subsection{Graph-Embedded Actor}
\subsubsection{Graph Modeling}
To characterize the interplay among neighboring agents, we model the THz reconfigurable massive MIMO-NOMA networks as a directed communication graph $\mathcal{G}=\left[\mathcal{I},\mathcal{E},\mathcal{T}_{\mathrm{N}},\mathcal{T}_{\mathrm{E}}\right]$,
where agents are modeled as two types of heterogeneous nodes $\mathcal{I}$, and interplay among agents are modeled as edges $\mathcal{E}$.
$\mathcal{T}_{\mathrm{N}}:\mathcal{I}\rightarrow\mathbb{C}^{d_{\mathrm{N}}^u}$ maps a node $i$ to its $d_{\mathrm{N}}^{u}$-dimensional node feature $\mathbf{o}_{i}^{u}(t)$.
Denote tuple $(i,i')$ as a directed edge from the source node $i$ to the destination node $i'$.
$\mathcal{T}_{\mathrm{E}}:\mathcal{E}\rightarrow\mathbb{C}^{d_{\mathrm{E}}^{ii'}}$ maps an edge $(i,i')$ to the $d_{\mathrm{E}}^{ii'}$-dimensional edge feature $\mathbf{d}_{ii'}(t)$.
Both $\mathbf{o}_i^{u}(t)$ and $\mathbf{d}_{ii'}(t)$ are extracted from the local observation of node $i\in\mathcal{I}^{u}$.

The node feature of a AP node $i \in \mathcal{M}$ contains the spatial channel information from AP $i$ to its associated users, the queue information of its associated users, and the local action-observation history of AP $i$, defined as
\begin{equation}\label{AP_nodefeature}
\mathbf{o}_{i}^{0}(t) = \left[\mathrm{vec}\left(\bm{\widetilde{\mathrm{H}}}^{ii}(t)\right),\bm{\Lambda}^{i}(t), \bm{\tau}_i^{0}(t) \right], \forall i \in \mathcal{I}^{0}.
\end{equation}
Moreover, the node feature of a RIS node $i \in \mathcal{I}^{1}$ includes the local action-observation history, i.e., $\mathbf{o}_{i}^{1}(t) = \bm{\tau}_i^{1}(t)$.

The edge feature $\mathbf{d}_{ii'}$ depicts the interplay effect of agent $i$ to $i'$, which can be mathematically denoted as
\begin{equation}\small
\begin{split}
&\mathbf{d}_{ii'}(t)=\\&
\begin{cases}
\left[\left\{\mathrm{vec}\left(\widetilde{\mathbf{H}}^{im}(t)\right)\right\}_{m\in\mathcal{N}_{1}^{i'}}\right], & i\in\mathcal{I}^{0},i'\in\mathcal{I}^{1}\text{, and } i'\in\mathcal{N}_{i}^{0},\\
\left[\mathrm{vec}\left(\mathbf{G}^{ii'}(t)\right),\mathrm{vec}\left(\mathbf{F}^{ii'}(t)\right)\right], & i\in\mathcal{I}^{1},i'\in\mathcal{I}^{0}\text{, and }i'\in\mathcal{N}_{j}^{1},\\
\mathrm{vec}\left(\bm{\widetilde{\mathrm{H}}}^{ii'}(t)\right), & i,i'\in\mathcal{I}^{0}\text{, and }i'\in\mathcal{N}_{i}^{0},\\
\mathbf{0}, & \text{otherwise}.
\end{cases}
\end{split}
\end{equation}
Specifically, for an AP-RIS link, the edge feature is the spatial channels from AP $i$ to the users served by neighboring APs of RIS $i'$.
For an AP-AP link, the edge feature denotes the channel from AP $i$ to the users associated with AP $i'$.
For a RIS-AP link, the edge feature includes the channels between RIS $i$ and both AP $i'$ and its associated users.

To learn the structured representation of the THz reconfigurable massive MIMO-NOMA networks, we integrate MPGNN \cite{MPGNN_Shen_2021} into the distributed actor/critic networks for embedding the graph.
Therefore, node/edge features of the directed graph over high-dimensional joint state space can be embedded into permutation-invariant features over low-dimensional space.
By exchanging the learned low-dimensional and permutation-invariant embedded features among neighboring agents, it's capable to improve generalization and enhance coordination among APs and RISs, while only requiring  low information exchange overhead.

\subsubsection{Graph-Embedded Actor}
The graph-embedded actor (policy) comprises a semi-distributed \textit{graph embedding module} which learns the embedding of the graph, followed by a distributed \textit{action generation module} that outputs action by taking the local embedded sate as inputs.
Here, we define $\pi_{\theta_{\mathrm{G}}}$ as the graph embedding sub-policy, and define $\pi_{\theta_{\mathrm{A}}}^{u}$, $\forall u$, as the action generation sub-policies, which are parameterized as MPGNNs and GRUs, respectively.
Let $\bm{\theta}_{\mathrm{G}}^{u}$ and $\bm{\theta}_{\mathrm{A}}^{u}$ denote the MPGNN and GRU weight parameters shared among agents $i \in \mathcal{I}^{u}$, respectively.

We maintain a MPGNN at each distributed node $i \in \mathcal{I}^{u}$.
Similar to the multi-layer perceptron (MLP), MPGNN exploits a layer-wise structure.
In each MPGNN layer, each agent first transmits embedded information to its neighbor agents,
and then aggregates embedded information from neighbor agents and updates its local hidden state.

Define the edges $(i,i')$ and $(i',i)$ as the outbound and inbound edges of node $i$.
Let $\mathcal{N}_{i+}^{u}$ and $\mathcal{N}_{i-}^{u}$ denote the sets of neighbor agents that are linked with agent $i$ through outbound and inbound edges of agent $i$, respectively.
In layer $n$, each agent $i\in\mathcal{I}^{u}$ locally embeds its node feature and the outbound edge feature $\mathbf{d}_{ii'}(t)$ as
\begin{equation}\label{local_embedding}
\mathbf{e}_{ii'}^{(n)}(t)=\psi_{u}^{(n)}\left(\mathbf{z}_{ui}^{(n-1)}(t),\mathbf{d}_{ii'}(t)\right), \forall i' \in \mathcal{N}_{i+}^{u}, i \in \mathcal{I}^{u},
\end{equation}
where $\mathbf{z}_{ui}^{(n-1)}(t)$ denotes the hidden state of MPGNN layer $n-1$ at agent $i \in \mathcal{I}^{u}$, and $\psi_{u}(\cdot)$ is the distributed embedding function.
Thereafter, each agent $i$ transmits the outbound embedded feature $\mathbf{e}_{ii'}^{(n)}(t)$ to its neighbor agents $i'\in\mathcal{N}_{i+}^{u}$,
and receives the inbound embedded feature $\mathbf{e}_{i'i}^{(n)}(t)$ from $i'\in\mathcal{N}_{i-}^{u}$.
Thereafter, the received embedded features are aggregated with a aggregation function $\phi_{u}(\cdot)$, which is a permutation-invariant function such as $sum(\cdot)$, $mean(\cdot)$ and $max(\cdot)$.
By combining the local hidden state and the aggregated features using the combining function $\Psi_{u}(\cdot)$, agent $i$ obtains the hidden state $\mathbf{z}_{ui}^{(n)}(t)$ of layer $n$ as
\begin{equation}\label{aggregation}
\mathbf{z}_{ui}^{(n)}(t)\!=\!\Psi_{u}^{(n)}\!\left(\!\mathbf{z}_{ui}^{(n-1)}(t),\phi_{u}^{(n)}\left(\!\left\{ \mathbf{e}_{i'i}^{(n)}\right\} _{i'\in\mathcal{N}_{i-}^{u}}(t)\!\right)\!\right)\!, \forall i \in \mathcal{I}^{u}.
\end{equation}
Combining \eqref{local_embedding} and \eqref{aggregation}, the update rule of the $n$-th MPGNN layer at node $i \in \mathcal{I}^{u}$ can be rewritten as
\begin{equation}\label{local_hidden_state}\small
\begin{split}
&\mathbf{z}_{ui}^{(n)}(t)\! = \\
&\!\Psi_{u}^{(n)}\!\left(\!\mathbf{z}_{ui}^{(n-1)}(t),\phi_{u}^{(n)}\!\left(\!\left\{ \psi_{u}^{(n)}\!\left(\!\mathbf{z}_{ui'}^{(n-1)}(t),\mathbf{d}_{i'i}(t)\!\right)\!\right\} _{i'\in\mathcal{N}_{i-}^{u}}\!\right)\!\right)\!,
\end{split}
\end{equation}
where $\mathbf{z}_{ui}^{(n)}(t)$ is the hidden/output state at $n$-th layer for agent $i\in\mathcal{I}^{u}$, and $\mathbf{z}_{ui}^{(0)}(t)=\mathbf{o}_{i}^{u}(t)$.

After the \textit{graph embedding module}, agent $i\in\mathcal{I}^{u}$ would predict local action utilizing GRU based on the output local embedded state $\widetilde{\mathbf{z}}_{i}^{u}(t)$, which is given by
\begin{equation}\label{embedding}
\widetilde{\mathbf{z}}_{i}^{u}(t) = \left[\mathbf{o}_{i}^{u}(t), \mathbf{z}_{ui}^{(N)}(t)\right].
\end{equation}

The local action $\mathbf{a}_{i}^{u}(t)$ for each agent $i\in\mathcal{I}^{u}$ is sampled from the action generation sub-policy  $\pi_{\theta_{\mathrm{A}}}^u\left(\mathbf{a}_{i}^{u}(t)|\widetilde{\mathbf{z}}_{i}^{u}(t)\right)$.

\begin{algorithm}[H]
\caption{Coordinated Multi-AP and Multi-RIS Scheduling Based on Distributed Graph-Embedded Actors}
\begin{algorithmic}[1]
\REQUIRE Distributed actors $\pi_{\theta}^u,\forall u \in \{0,1\}$.
\STATE Initialize $t=1$, $q_{nk}^{m}(t) = 0$, $Y_{nk}^{m}(t)=0$, $\forall k \in \mathcal{K}$;
\FOR{$t = 1,2,...,T$}
    \STATE Each agent obtains local observation as \eqref{local_obs_AP} and \eqref{local_obs_RIS};
    \FOR{each MPGNN layer $n\in\{1,2,...,N\}$}
        \STATE Each agent $i\in\mathcal{I}^{u}$ calculates local feature embedding $\mathbf{e}_{ii'}^{(n)}(t), \forall i' \in \mathcal{N}_{i+}^{u}$, using \eqref{local_embedding};
        \STATE Each agent $i\!\in\!\mathcal{I}^{u}$ aggregates feature embedding $\mathbf{e}_{i'i}^{(n)}(t)$ from neighbor agents $i' \!\in\! \mathcal{N}_{i-}^{u}$, and update the hidden state $\mathbf{z}_{ui}^{(n)}(t)$ based on \eqref{aggregation};
    \ENDFOR
    \STATE Each agent $i \in \mathcal{I}^{u}$ samples local action $\mathbf{a}_{i}^{u}(t)$ according to $\pi_{\theta_{\mathrm{A}}}^u\left(\mathbf{a}_{i}^{u}(t)|\widetilde{\mathbf{z}}_{i}^{u}(t)\right)$;
    \STATE Each AP $m$ calculates QoS-based user clustering, obtains hybrid beamformer $\mathbf{V}^{m}(t)$ and $\mathbf{W}^m(t)$, and decides decoding order of IoT users for each user cluster;
    \STATE Observe the global reward $r\left(\mathbf{s}(t),\mathbf{a}(t)\right)$, and calculate $q_{nk}^{m}(t+1)$ and $Y_{nk}^{m}(t+1)$ according to \eqref{queue} and \eqref{virtual_queue};
\ENDFOR
\ENSURE $\left\{\mathbf{s}(t),\widetilde{\mathbf{z}}(t),\mathbf{a}(t),r\left(\mathbf{a}(t),\mathbf{s}(t)\right),\mathbf{s}(t+1)\right\}_{t\in\{1,2,...,T\}}$
\end{algorithmic}
\label{alg:actor}
\end{algorithm}

The coordinated scheduling for APs and RISs based on the distributed graph-embedded actors can be summarized as Algorithm \ref{alg:actor}.
At each time slot $t$, APs and RISs exchange embedded features and obtain their local embedded state $\widetilde{\mathbf{z}}_{i}^{u}$ through MPGNN.
Thereafter, each agent $i \in \mathcal{I}^{u}$ samples its local action based on the action generation sub-policy $\pi_{\theta_{\mathrm{A}}}^{u}\left(\mathbf{a}_{i}^{u}(t)|\widetilde{\mathbf{z}}_{i}^{u}(t)\right)$ of the local actor.
By observing the reconfigured channels, each AP further decides the user clustering, hybrid beamforming and IoT user decoding order.
In this way, the customized user clustering, NOMA decoding and hybrid beamforming schemes are considered as part of the environment feedback, which would be implicitly learned by the distributed actors through exploration and exploitation.
The experiences are then joined into the replay memory buffer $\mathcal{B}$ for centralized learning.

\subsubsection{Policy improvement}
Denote the combined parameters of \textit{graph embedding module} and \textit{action generation module} in the distributed actor (policy) as $\bm{\theta}_{\pi}^{u} = \left[ \bm{\theta}_{\mathrm{G}}^{u}, \bm{\theta}_{\mathrm{A}}^{u} \right]$.
Here, the distributed actors $\bm{\theta}_{\pi} = [\{\bm{\theta}_{\pi}^{u}\}_{u\in\{0,1\}}]$ are trained to maximize the following performance function:
\begin{equation}\label{policy_loss}
L_{\pi}(\bm{\theta}_{\pi}^{u}) = \mathbb{E}_{\mathbf{s}\sim p^{\pi},\mathbf{a}\sim\pi}\left[r(\mathbf{s}(t),\mathbf{a}(t))\right],
\end{equation}
where $p^{\pi}$ is the joint state transition by following the joint policy $\pi$.
Therefore, we calculate policy gradient based on the advantage function, which is given by
\begin{equation}\label{policy_gradient}
\Delta \bm{\theta}_{\pi}^{u} (t) = \sum_{i} \nabla_{\bm{\theta}_{\pi}^{u}}\log
\pi^{u}\left(\mathbf{a}_{i}^{u}(t)|\hat{\mathbf{s}}_{i}^{u}(t)\right)
A \left(\mathbf{s}(t),\mathbf{a}(t)\right),
\end{equation}
where $ \hat{\mathbf{s}}_{i}^{u}(t)$
 is the actual input of the graph-embedded actor.
Moreover, $A\left(\mathbf{s}(t),\mathbf{a}(t)\right)$ denotes the temporal difference (TD) advantage, which is given by
\begin{equation}\label{Advantage}
\begin{split}
A\left(\mathbf{s}(t),\mathbf{a}(t)\right) &= Q(\mathbf{s}(t),\mathbf{a}(t)) - V_{tot}(\mathbf{s}_{t}) \\
&= r(\mathbf{s}(t),\mathbf{a}(t))+\Gamma V_{tot}(\mathbf{s}_{t+1}) - V_{tot}(\mathbf{s}_{t}).
\end{split}
\end{equation}
Here, $V_{tot}(\mathbf{s}_{t})$ and $Q(\mathbf{s}(t),\mathbf{a}(t))$ denote the global state value and the global state-action value, respectively.

\subsection{Value Decomposition}
To address credit assignment problem during training, we rely on the value-decomposition critics \cite{VDAC_Su_2021} to train the distributed actors, which decomposes the global state value $V_{\mathrm{tot}}(\mathbf{s}(t))$ into local critics that are combined with a mixing function $f^{\mathrm{mix}}$ as
\begin{equation}\small
\begin{split}
V_{\mathrm{tot}}(\mathbf{s}(t)) =&
f^{\mathrm{mix}}\bigg(\mathbf{s}(t), V^{0}\left(\widetilde{\mathbf{z}}_{1}^{0}(t)\right), V^{0}\left(\widetilde{\mathbf{z}}_{2}^{0}(t)\right),..., V^{0}\left(\widetilde{\mathbf{z}}_{M}^{0}(t)\right), \\&
V^{1}\left(\widetilde{\mathbf{z}}_{1}^{1}(t)\right), V^{1}\left(\widetilde{\mathbf{z}}_{2}^{1}(t)\right), ...,V^{1}\left(\widetilde{\mathbf{z}}_{J}^{1}(t)\right)\bigg),
\end{split}
\end{equation}
where $V^{u}\left(\widetilde{\mathbf{z}}_{i}^{u}(t)\right)$ is the local state value for agent $i\in\mathcal{I}^{u}$.

Now, we are ready to introduce the following Lemma.
\begin{lemma}\label{Lemma_VD}
Any action $\mathbf{a}_{i}^{u}$ that can improve agent $i$'s local critic value at local embedded state $\widetilde{\mathbf{z}}_{i}^{u}$ will also improve the global state value $V_{\mathrm{tot}}$,
if the other agents stay at the same local embedded states taking actions $\mathbf{a}_{-i}^{u}$, and the following relationship holds
\begin{equation}\label{V_partial_derivation}
\frac{\partial V_{\mathrm{tot}}\left(\mathbf{s}(t)\right)}{\partial V^{u}\left(\widetilde{\mathbf{z}}_{i}^{u}(t)\right)} \geq 0, \forall i\in\mathcal{I}^{u}, u\in\{0,1\}.
\end{equation}
\begin{proof}
Based on \eqref{V_partial_derivation}, the global reward monotonically increase with $V^{u}\left(\widetilde{\mathbf{z}}_{i}^{u}(t)\right)$ when the other agents stay at the same local embedded states taking actions $\mathbf{a}_{-i}^{u}(t)$.
Therefore, if a local action $\mathbf{a}_{i}^{u}(t)$ is capable to improve $V^{u}\left(\widetilde{\mathbf{z}}_{i}^{u}(t)\right)$, obviously it can also improve $V_{\mathrm{tot}}\left(\mathbf{s}(t)\right)$.
\end{proof}
\end{lemma}

\begin{remark}
Based on Lemma \ref{Lemma_VD}, it's rational to approximately decompose the high-complexity global critic into distributed critics combined by a mixing network with non-negative network weight parameters, thus reducing complexity and achieving \textit{difference rewards}.
\end{remark}

During centralized training, each agent attains a differential reward $V^{u}\left(\widetilde{\mathbf{z}}_{i}^{u}(t)\right)$ based on the local graph-embedded features to evaluate its contribution for global reward improvement, which can further facilitate agents' coordination.
The weights of mixing network are generated through separate hypernetworks \cite{hypernet_Ha_2016}.
Each hypernetwork takes the joint state and state-value as input to compute the weights of one layer of the mixing network \cite{QMIX_Rashid_2018}.
To guarantee that the output weights are non-negative, the absolute activation function is exploited at the hypernetworks.

Define $\bm{\theta}_{\mathrm{V}}^{u}$ as the weight parameters of distributed critic $V_{\theta}^{u}$ that are shared among agents $i \in \mathcal{I}^{u}$, and denote $\bm{\mu}$ as the weights of the mixing network $f_{\mu}^{\mathrm{mix}}$.
The distributed critic and mixing network are optimized by mini-batch gradient descent to minimize the following critic loss:
\begin{equation}\label{Critic_Loss}
\begin{split}
& L_{V}\left(\bm{\theta}_{\mathrm{V}}^{u}(t), \bm{\mu}(t)\right)
= \left[\hat{r}_t\left(\mathbf{s}(t),\mathbf{a}(t)\right) -V_{\mathrm{tot}}\left(\mathbf{s}(t)\right)\right]^{2} \\
\! &= \! \bigg[ \hat{r}_t\left(\mathbf{s}(t),\mathbf{a}(t)\right) - \!
\! f_{\mu}^{\mathrm{mix}} \!\bigg(\! \mathbf{s}(t), \!
V_{\theta}^{0}\left(\widetilde{\mathbf{z}}_{1}^{0}(t)\right), \! ..., \!
V_{\theta}^{1}\left(\widetilde{\mathbf{z}}_{J}^{1}(t)\right) \!\bigg) \!\bigg]^{2} \!,
\end{split}
\end{equation}
where
$\hat{r}_t\left(\mathbf{s}(t),\mathbf{a}(t)\right)=\sum_{i=1}^{n}\Gamma^{i-1}r_{t+i}+\Gamma^{n}V_{\mathrm{tot}}\left(\mathbf{s}_{t+n}\right)$ is the $n$-step return bootstrapped from the last state, with $n$ upper-bounded by $T$.
Therefore, the mixing networks can be updated by
\begin{equation}\label{Mixing_Update}
\bm{\mu} \leftarrow \bm{\mu} -\kappa_{\mu}\sum_{t}\nabla_{\bm{\mu}} L_{\mathrm{V}}\left(\bm{\theta}_{\mathrm{V}}^{u}(t), \bm{\mu}(t)\right),
\end{equation}
where $\kappa_{\mu}$ is the learning rate for mixing network update.

To reduce complexity, we further share the weight parameters of non-output layers between distributed critics and actors for agents of the same type, similar to \cite{VDAC_Su_2021}.
Denote the combined weight parameters of the distributed actors and critics networks as $\bm{\theta}^{u} = \left[\bm{\theta}_{\mathrm{G}}^{u},  \bm{\theta}_{\mathrm{A}}^{u}, \bm{\theta}_{\mathrm{V}}^{u}\right]$.
Therefore, the critic gradient with respect to $\bm{\theta}^{u}$ can be given by
\begin{equation}\label{Critic_Gradient}
\Delta \bm{\theta}_{\mathrm{V}}^{u}(t) = \nabla_{\bm{\theta}^{u}} L_{\mathrm{V}}\left(\bm{\theta}_{\mathrm{V}}^{u}(t), \bm{\mu}(t)\right).
\end{equation}
Hence, the update rule of the distributed actor/critic networks can be derived as
\begin{equation}\label{Theta_Update}
\bm{\theta}^{u} \leftarrow \bm{\theta}^{u} +
\sum_{t} \left( \kappa_{\pi} \Delta \bm{\theta}_{\pi}^{u}(t)
- \kappa_{\mathrm{V}} \Delta \bm{\theta}_{\mathrm{V}}^{u}(t)\right),
\end{equation}
where $\kappa_{\pi}$ and $\kappa_{\mathrm{V}}$ are the learning rate for policy improvement and critic learning, respectively.

The whole \textit{GE-VDAC} based MADRL algorithm is summarized in Algorithm \ref{alg:seq}.

\begin{algorithm}[H]
\caption{\textit{GE-VDAC} Based MADRL Algorithm}
\begin{algorithmic}[1]
\STATE Initialize the parameters of distributed actors/critics $\bm{\theta}^{u}, \forall u$, and the mixing network $\bm{\mu}$;
\STATE Initialize the learning rates $\kappa_{\mu}, \kappa_{\pi}, \kappa_{\mathrm{V}}$;
\FOR{each training episode}
    \STATE Initialize replay buffer $\mathcal{B} = \emptyset$;
    \FOR{each semi-distributed execution stage}
        \STATE Execute the semi-distributed multi-AP multi-RIS scheduling for $T$ steps based on algorithm \ref{alg:actor};
        \STATE Add experiences $\big\{\mathbf{s}(t),\widetilde{\mathbf{z}}(t),\mathbf{a}(t),r\left(\mathbf{a}(t),\mathbf{s}(t)\right),\mathbf{s}(t+1)\big\}_{t\in\{1,2,...,T\}}$ to replay buffer $\mathcal{B}$;
    \ENDFOR
    \STATE Draw a batch of experiences from buffer $\mathcal{B}$;
    \STATE Calculate the policy gradient $\bm{\theta}_{\pi}^{u},\forall u$, based on \eqref{policy_gradient};
    \STATE Calculate the critic loss based on \eqref{Critic_Loss};
    \STATE Update mixing network weights $\bm{\mu}$ based on \eqref{Mixing_Update};
    \STATE Calculate the critic gradients $\Delta \bm{\theta}_{\mathrm{V}}^{u},\forall u$, based on \eqref{Critic_Gradient};
    \STATE Update distributed graph-embedded actors and critics based on \eqref{Theta_Update}.
\ENDFOR
\end{algorithmic}
\label{alg:seq}
\end{algorithm}

\subsection{Theoretical Analysis of GE-VDAC}
\subsubsection{Permutation invariance}
Firstly, we show that \textit{GE-VDAC} has permutation-invariant property, which can lead to better generalization.
Consider a graph $\mathcal{G}$, whose node and edge features are denoted by $\mathbf{O}$ and $\mathbf{D}$.
Let $\nu$ denote the permutation operator.
Let $\nu\star \mathcal{I}$ and $\nu\star \left(\mathbf{O}, \mathbf{D}\right)$ denote the permutation of nodes (agents) and graph features, respectively.
Given two graphs $\mathcal{G}$ and $\mathcal{G}'$, if their exists a permutation $\nu$ satisfying $\left(\mathbf{O}, \mathbf{D}\right) = \nu\star\left(\mathbf{O}', \mathbf{D}'\right)$,
then $\mathcal{G}$ and $\mathcal{G}'$ are \textit{isomorphic}, denoted as $\mathcal{G}\cong\mathcal{G}'$.
\begin{definition}
Given two \textit{isomorphic} communication graphs $\mathcal{G}\cong\mathcal{G}'$ with permutation $\left(\mathbf{O}, \mathbf{D}\right) = \nu\star\left(\mathbf{O}', \mathbf{D}'\right)$, the distributed actors/critics are permutation invariant if the output
joint action vector $\mathbf{a},\mathbf{a}'$ and joint critic vector $\mathbf{v},\mathbf{v}'$ satisfy
\begin{equation}
\mathbf{a} =\nu\star\mathbf{a}', \mathbf{v} = \nu\star \mathbf{v}'.
\end{equation}
\end{definition}

Traditional centralized/multi-agent actor-critic algorithms usually exchange information directly and utilize neural networks such as MLP, convolutional neural network (CNN) and recurrent neural network (RNN) to learn actor and critic, which cannot achieve the permutation invariance property.
Despite a sample and all of its agents' permutations actually represent the same environments, they would be viewed as utterly different samples.
Since traditional algorithms cannot adapt to agents' permutations intelligently, all permutations of one sample should be fed to train actors and critics, leading to poor generalization and learning speed.
However, by extending the traditional DRL algorithms, the graph embedding based distributed actors and critics in \textit{GE-VDAC} enjoy the permutation invariance property, as shown below.

\begin{proposition}\label{Proposition_Permutation_Invariance}
If the aggregation functions $\phi_u(\cdot), \forall u$, utilized in \eqref{aggregation} and \eqref{local_hidden_state} are permutation invariant, the distributed actors/critics based on \textit{GE-VDAC} satisfy permutation invariance.
\begin{proof}
Considering the update rule of hidden states in \eqref{local_hidden_state}, if aggregation functions $\phi_u(\cdot), \forall u$, are permutation invariant,
the graph embedding features obtained by  MPGNN $\pi_{\theta_{\mathrm{G}}}^{u}, \forall u$, enjoy permutation invariant property \cite{MPGNN_Shen_2021}, i.e., $\pi_{\theta_{\mathrm{G}}}^{u}\left(\nu\star\mathbf{O},\nu\star\mathbf{D}\right)
= \nu\star\left(\pi_{\theta_{\mathrm{G}}}^{u}\left(\mathbf{O},\mathbf{D}\right)\right)$.
Given agents' permutation $\left(\mathbf{O}',\mathbf{D}'\right)=\left(\nu\star\mathbf{O},\nu\star\mathbf{D}\right)$, we have
\begin{equation}\label{Feature_Permutation_Invariant}
\left(\mathbf{z}^{u}\right)'
=\pi_{\theta_{\mathrm{G}}}^{u}\left(\nu\star\mathbf{O},\nu\star\mathbf{D}\right)
=\nu\star\left(\pi_{\theta_{\mathrm{G}}}^{u}\left(\mathbf{O},\mathbf{D}\right)\right)=\nu\star\mathbf{z}^{u}, \forall u.
\end{equation}
Therefore, we can obtain $\left(\widetilde{\mathbf{z}}_{i}^{u}\right)' = \widetilde{\mathbf{z}}_{\nu(i)}^{u}, \forall i, \forall u$, which implies the permutation equivalence of the local embedded states.
Considering the definition of the local critics, we can achieve
\begin{equation}
\mathbf{v}'=\left\{ V_{\theta}^{u}\left(\left(\widetilde{\mathbf{z}}_{i}^{u}\right)'\right)\right\} _{i\in\left(\nu\star\mathcal{I}\right)}
=\left\{ V_{\theta}^{u}\left(\widetilde{\mathbf{z}}_{\nu(i)}^{u}\right)\right\} _{i\in\left(\nu\star\mathcal{I}\right)}
=\nu\star\mathbf{v}.
\end{equation}
Moreover, since the local actions are sampled from
$\mathbf{a}_{\nu(i)}^{u}\sim\pi_{\theta_{\mathrm{A}}}^{u}\big(\cdot|\widetilde{\mathbf{z}}_{\nu(i)}^{u}\big)$ and
$\left(\mathbf{a}_{i}^{u}\right)'\sim\pi_{\theta_{\mathrm{A}}}^{u}\big(\cdot|\left(\widetilde{\mathbf{z}}_{i}^{u}\right)'\big)$,
we have
\begin{equation}
\!\pi_{\theta_{\mathrm{A}}}^{u}\!\left(\left(\mathbf{a}_{i}^{u}\right)'|\left(\widetilde{\mathbf{z}}_{i}^{u}\right)'\right)\!
\!=\! \pi_{\theta_{\mathrm{A}}}^{u}\big(\mathbf{a}_{\nu(i)}^{u}|\widetilde{\mathbf{z}}_{\nu(i)}^{u}\big),
\forall i \!\in\! \mathcal{I}^{u}, u\!\in\!\{0,\!1\},
\end{equation}
which signifies $\mathbf{a}'=\nu\star\mathbf{a}$.
This ends the proof.
\end{proof}
\end{proposition}

Based on Proposition \ref{Proposition_Permutation_Invariance}, the learned distributed actors/critics in \textit{GE-VDAC} are insusceptible to the agents' permutations.
Therefore, \textit{GE-VDAC} can improve the generalization and the learning speed in multi-agent cooperative learning.

\subsubsection{Convergence Guarantee}
We also demonstrate that the proposed \textit{GE-VDAC} algorithm can achieve the locally optimal policy.
The convergence of the \textit{GE-VDAC} algorithm can be theoretically guaranteed with the following theorem.
\begin{theorem}\label{Convergence}
The proposed algorithm can converge to a locally optimal policy for the Dec-POMDP, i.e.,
\begin{equation}\label{optimum}
\lim \inf_t \|\nabla_{\bm{\theta}_{\pi}} L_{\pi}(\bm{\theta}_{\pi})\|=0, w.p.1.
\end{equation}
\begin{proof}
See Appendix \ref{Proof_Convergence}.
\end{proof}
\end{theorem}

\subsubsection{Complexity Analysis}

The computational complexity of the proposed \textit{GE-VDAC} algorithm mainly comes from MPGNNs and GRUs.
For each MPGNN, the local embedding function $\psi(\cdot)$ and the combining function $\Psi(\cdot)$ are parameterized as two-layer MLPs with $S_{\psi}$ and $S_{\Psi}$ neurons in each MLP layer, respectively.
Thus, the computational complexity of local embedding, aggregation, and combination at agent $i\in\mathcal{I}^{u}$ in MPGNN layer $n$ can be respectively written as $ \mathcal{O}\!\left(\mathrm{deg}_{i+}^{u}\left(S_{\mathrm{Z}}^{(n-1)}S_{\psi}\!+\!S_{\mathrm{E}}S_{\psi}\right)\right)$, $\mathcal{O}\!\left(\mathrm{deg}_{i-}^{u}S_{\mathrm{E}}\right)$, and
$\mathcal{O}\!\left(S_{\Psi}\left(S_{\mathrm{Z}}^{(n-1)}\!+\!S_{\mathrm{E}}\right)\!+\!S_{\Psi}S_{\mathrm{Z}}^{(n)}\right)$,
where the outdegree/indegree $\mathrm{deg}_{i\pm}^{u}=\left|\mathcal{N}_{i\pm}^{u}\right|$ indicates the number of outbound/inbound edges of agent $i\in\mathcal{I}^{u}$,
$S_{\mathrm{E}}$ is the size of the local embedding $\mathbf{e}_{ii'}^{(n)}$,
and $S_{\mathrm{Z}}^{(n)}$ represents the size of the output vector at layer $n$ that is initialized as $S_{\mathrm{Z}}^{(0)}=\left(d_{\mathrm{N}}^{u}+\mathrm{d}_{\mathrm{E}}^{ii'}\right)$.
Since all the agents can be calculated in parallel, the computational complexity of the whole $N$ MPGNN layers is $\mathcal{O}\left(\sum_{n=1}^{N}C_{\max}^{(n)}\right)$,
where $C_{\max}^{(n)}$ implies the maximal complexity of agents in MPGNN layer $n$, i.e.,
\begin{multline}\label{complexity_MPGNN}
C_{\max}^{(n)} = \max_{i \in \mathcal{I}^{u}, \atop u\in\{0,1\}} \bigg[ \mathrm{deg}_{i+}^{u} S_{\psi}\left(N_{\mathrm{Z}}^{(n-1)}+S_{\mathrm{E}}\right)\\
 + \mathrm{deg}_{i-}^{u}S_{\mathrm{E}}
+ S_{\Psi}\left(S_{\mathrm{Z}}^{(n-1)}+S_{\mathrm{E}}+S_{\mathrm{Z}}^{(n)}\right) \bigg].
\end{multline}
On the other hand, since the complexity of GRU per weight and time step is $\mathcal{O}(1)$, its computational complexity depends on the number of GRU weight parameters \cite{LSTM_complexity_2020}.
For agent $i\in\mathcal{I}^{u}$, the total parameter number of one GRU cell is $3\left(S_{\mathrm{in}}^{u}+S_{\mathrm{GRU}}\right)S_{\mathrm{GRU}}$, where
$S_{\mathrm{in}}^{u} =  S_{\mathrm{Z}}^{(N)}+d_{\mathrm{N}}^{u}$ denotes the size of the GRU input defined in \eqref{embedding},
and $S_{\mathrm{GRU}}$ is the number of neurons for each gate in the GRU cell.
Based on the above analyses, the computational complexity of the proposed \textit{GE-VDAC} algorithm can be obtained by
\begin{equation}\label{complexity}
\mathcal{O}\left(\sum_{n=1}^{N}C_{\max}^{(n)}+3\left(S_{\mathrm{Z}}^{(N)}+\max_{u\in\{0,1\}}d_{\mathrm{N}}^{u}+S_{\mathrm{GRU}}\right)S_{\mathrm{GRU}}\right).
\end{equation}

In comparison, the computational complexity of the traditional \emph{VDAC} algorithm \cite{VDAC_Su_2021} (without information interaction) can be obtained by modifying the GRU input size and ignoring the MPGNN part in the \textit{GE-VDAC} algorithm as
$\mathcal{O}\left(3\left[S_{\mathrm{in}}^{\mathrm{VDAC}}+S_{\mathrm{GRU}}\right]S_{\mathrm{GRU}}\right)$,
where $S_{\mathrm{in}}^{\mathrm{VDAC}} = \max_{{i\in\mathcal{I}^{u},\atop u\in\{0,1\}}}\left(d_{\mathrm{N}}^{u}+\sum_{i'\in\mathcal{N}_{i+}^{u}}d_{\mathrm{E}}^{ii'}\right)$ is the maximal size of the GRU input at all agents.  On the other hand, an advanced \emph{VDAC} algorithm with direct information exchange among agents, denoted as \emph{IE-VDAC}, has the complexity of $\mathcal{O}\left(3\left[S_{\mathrm{in}}^{\mathrm{IE-VDAC}}+S_{\mathrm{GRU}}\right]S_{\mathrm{GRU}}\right)$, where $S_{\mathrm{in}}^{\mathrm{IE-VDAC}} = \max_{{i\in\mathcal{I}^{u},\atop u\in\{0,1\}}}\left(d_{\mathrm{N}}^{u}+\sum_{i'\in\mathcal{N}_{i+}^{u}}d_{\mathrm{E}}^{ii'}+\sum_{i'\in\mathcal{N}_{i-}^{u}}d_{\mathrm{E}}^{i'i}\right)$.

\section{Numerical Results}

In this section, we present numerical results to demonstrate the effectiveness of the proposed smart reconfigurable massive MIMO-NOMA networks.
We consider a network with $M=3$ APs uniformly deployed on the ceiling of an $8\times5$ m$^2$ indoor room. Furthermore, we respectively consider $J=\{1, 2, 4\}$ RISs mounted on the walls. When $J=1$, RIS is deployed in the middle of one wall. When $J\in\{2,4\}$, RISs are symmetrically and uniformly spaced on walls, respectively.
Each AP $m$ equips $N_{\mathrm{R}}=4$ RF chains to serve $K_{\mathrm{S}} = 4$ SE users and $K_{\mathrm{U}} = K^m - K_{\mathrm{S}}$ IoT users.
For simplicity, we assume that the numbers of users associated with each AP are the same.
The minimum rate requirements for SE users and IoT users are $2$ Gbit/s and $0.1$ Gbit/s, respectively.
We further assume both SE and IoT users arrive the network under the Poisson distribution with the density of $10$ Gbit/s and $0.2$ Gbit/s, respectively.
Moreover, the maximum queue lengths for SE users and IoT users are set as $25$ Gbit/s and $10$ Gbit/s, and the violation probability is limited by $0.1$.
On the other hand, each RIS equips with $L=20$ cost-effective RIS elements with quantization bit $B^{\mathrm{R}}=\{1,2\}$.
The LoS probabilities between AP-device and RIS-device links are calculated according to the human blockage model in \cite{Blocakge_Wu_2021}.
Moreover, the reflection coefficient of NLoS path is given by $\lambda = \rho^{\mathrm{F}} \rho^{\mathrm{R}}$ \cite{NLoSReflection_Piesiewicz_2007},
where $\rho^{\mathrm{F}} = \frac{\cos\varphi_{\mathrm{in}} - \sqrt{n_{\mathrm{r}}^{2}-\sin^2\varphi_{\mathrm{in}}}}{\cos\varphi_{\mathrm{in}} + \sqrt{n_{\mathrm{r}}^{2}-\sin^2\varphi_{\mathrm{in}}}}$ denotes the Fresnel reflection coefficient, and $\rho^{\mathrm{R}} = e^{-\frac{1}{2}\left(\frac{4\pi f\sigma_{\mathrm{s}}\cos\varphi_{\mathrm{in}}}{c}\right)}$ is the Rayleigh roughness factor.
$\varphi_{\mathrm{in}}$ represents the angle of incidence and reflection, $n_{\mathrm{r}} = 1.922 + 0.0057j$ means the refractive index, and $\sigma_{\mathrm{s}}=0.05$e-3 denotes the standard deviation of the surface characterizing the material roughness.
Furthermore, the AWGN power spectral density is $\sigma^2=-174$ dBm/Hz.
The simulation parameters and the neural network (NN) settings are summarized in Table \ref{table:param}.
For \emph{GE-VDAC}, we empirically learn $N=2$ MPGNN layers with local embedding size $S_{\mathrm{E}}=32$ and output size $S_{\mathrm{Z}}^{(1)}=S_{\mathrm{Z}}^{(2)}=48$.

\begin{table}[t]
	\begin{minipage}{1\linewidth}
		\centering
		\caption{Simulation Parameters}
		\label{table:param}
		\resizebox{1\textwidth}{!}{
			\begin{tabular}{lc|cc}
				\hline
				Parameters & Values & NN parameters & Values \\ \hline
				Frequency         & $f$ = $0.3$ THz      & GNN layers          & $N=2$ \\
                Absorption coeff. & $k(f)=0.0033$ $m^{-1}$ & Local embedding     & $S_{\mathrm{E}} = 32$\\
				Bandwidth 	      & $10$ GHz             & MLP neurons         & $S_{\psi} = S_{\Psi} = 48$\\
                Antenna gain      & $\varUpsilon=20$ dBi & Combining function  & $\phi_u = mean(\cdot)$\\
                Time slots        & $T=40$               & MPGNN outputs        & $S_{\mathrm{Z}}^{(1)} = S_{\mathrm{Z}}^{(2)} = 48$ \\
                Inefficiency      & $\xi=1/0.38$         & 1st FC layer   & $\left(S_{\mathrm{in}}^{u}, S_{\mathrm{GRU}}\right)$\\
                Maximum power     & $P_{\mathrm{AP}}^{\max}=5$ W& GRU neurons   & $S_{\mathrm{GRU}} = 64$\\
                Baseband power    & $P_{\mathrm{BB}}=0.2$ W& Last FC layer & $\left(S_{\mathrm{GRU}}, \text{action size}\right)$ \\
                RF chain power & $P_{\mathrm{RF}}=0.16$ W &   \multirow{2}*{\makecell[c]{Learning rates ($\kappa_{\mu},$\\$\kappa_{\pi}$, and $\kappa_{\mathrm{V}}$)}} &\multirow{2}*{$0.0005$} \\
                Phase shifter power & $P_{\mathrm{PS}}=0.03$ W & ~ & ~ \\
                Amplifier power  & $P_{\mathrm{PA}}=0.02$ W\\
                RIS element power & $P_{\mathrm{RIS}}=B^{\mathrm{R}}\times0.01$ W&           &\\ \hline
			\end{tabular}
		}
	\end{minipage}
\hfill
\end{table}

\begin{figure}[htbp]
\centering
\subfloat[Convergence of the test reward.]{\label{rewards}
    \includegraphics[width=.48\textwidth]{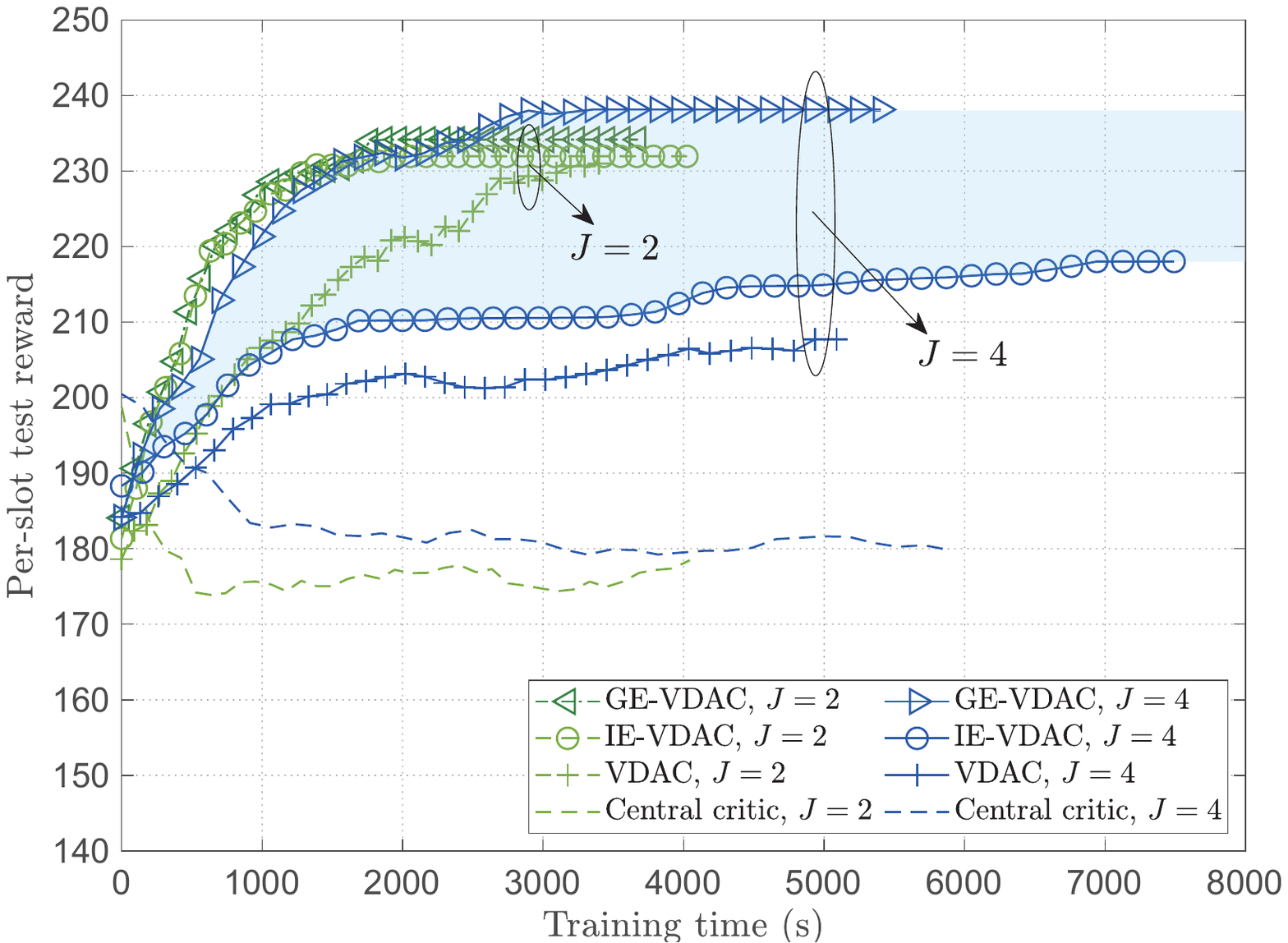}
}
\\
\subfloat[Convergence of the energy efficiency.]{\label{EE_comparison}
    \includegraphics[width=.48\textwidth]{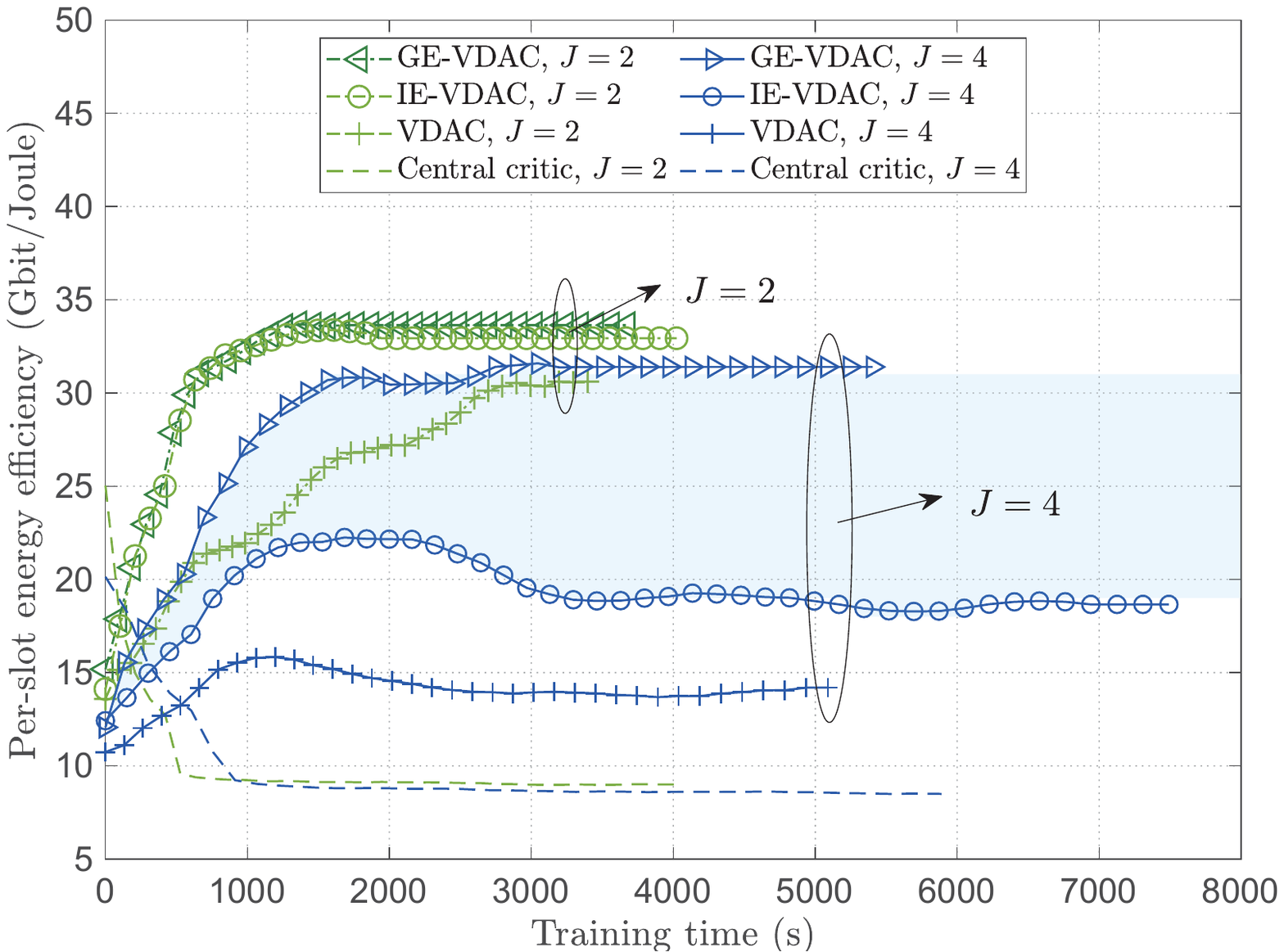}
}
\caption{Convergence performance comparisons among different algorithms.}
\label{fig_convergence}
\end{figure}

\begin{figure}[htbp]
\centering
\subfloat[Energy efficiency.]{\label{EE_K}
    \includegraphics[width=.48\textwidth]{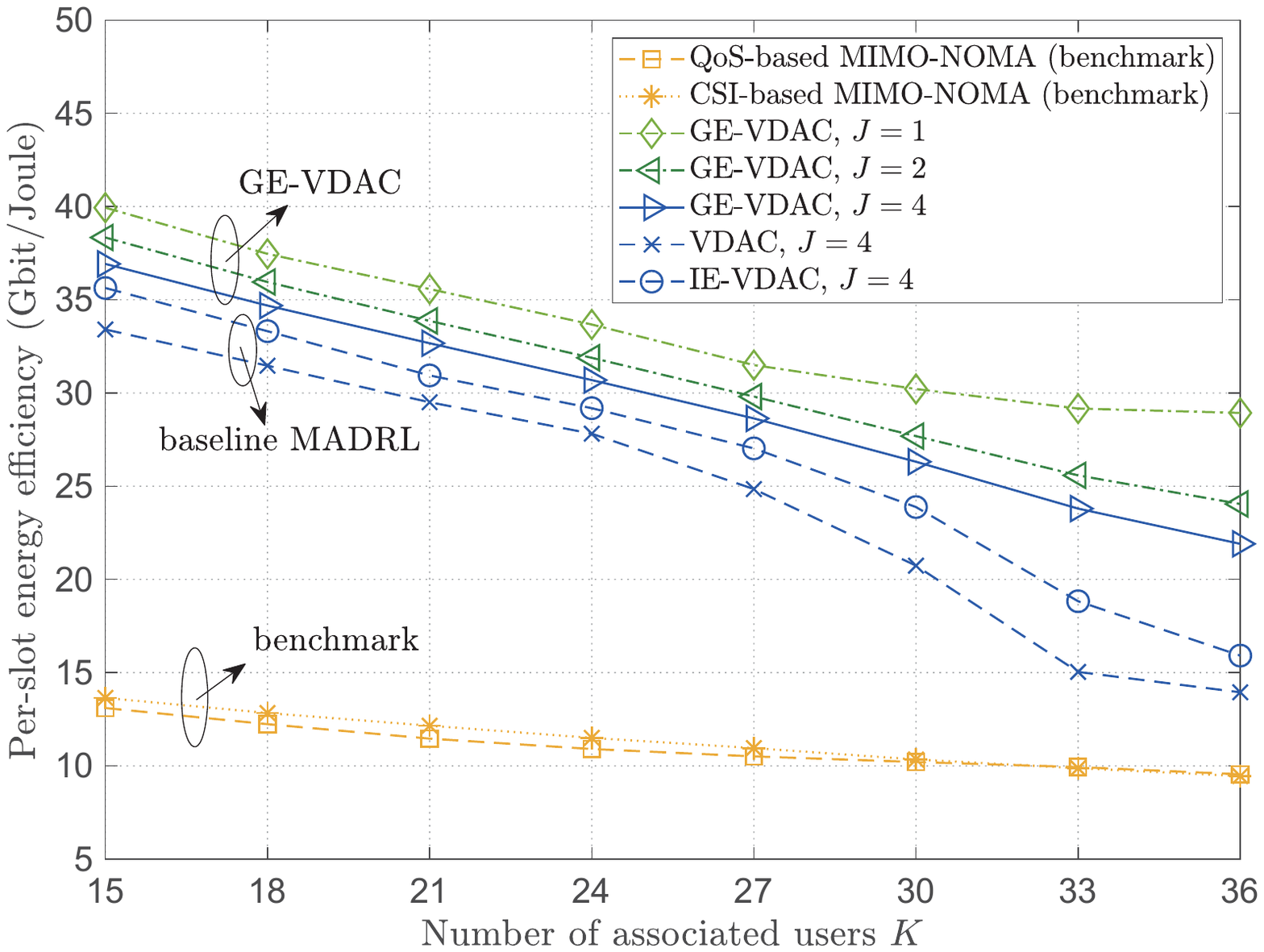}
}
\\
\subfloat[Reliability of SE users.]{\label{Relia_K}
    \includegraphics[width=.48\textwidth]{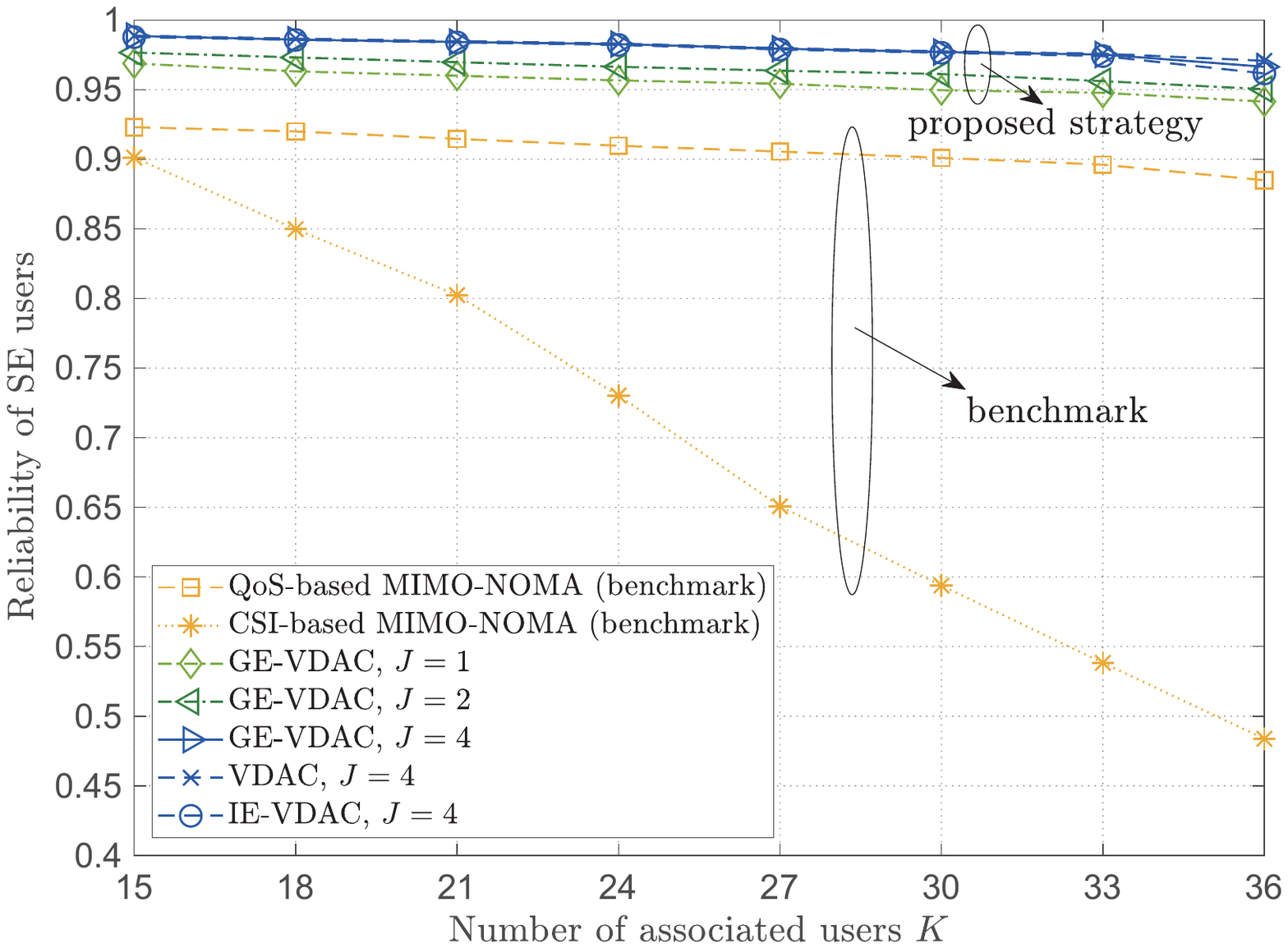}
}
\caption{Energy efficiency and reliability performance under different algorithms.}
\label{fig_K}
\end{figure}


We compare the proposed \textit{GE-VDAC} algorithm with the following three baselines:
\begin{itemize}
  \item \textbf{Central critic}: the multi-agent centralized-critic distributed-actor algorithm, which extends the traditional multi-agent actor-critic algorithm \cite{MAAC_Ryan_2017} to the hybrid discrete and continuous action space.
  \item \textbf{\textit{VDAC}}: the fully distributed \textit{VDAC} algorithm \cite{VDAC_Su_2021} without information exchange among agents.
  \item \textbf{\textit{IE-VDAC}}: the \textit{VDAC} algorithm with direct \textbf{I}nformation \textbf{E}xchange among neighboring agents.
\end{itemize}
The convergence performance comparisons among different algorithms are shown in Fig. \ref{fig_convergence}, where all of the agents are trained for $150000$ time steps, and we set $K = 24$, $N_{\mathrm{A}} = 64$, $b=1$.
We can see that the value-decomposition based algorithms (\textit{VDAC, IE-VDAC, GE-VDAC}) outperform central critic algorithm, since the coordination among agents is enhanced through \textit{difference rewards}.
The fully distributed \emph{VDAC} algorithm requires no information exchange overhead and takes the least training time, but has lower learning speed and expected reward than both \textit{IE-VDAC} and \textit{GE-VDAC}.
With less information exchange overhead than \textit{IE-VDAC}, \textit{GE-VDAC} can achieve the highest expected reward and learning speed, and the performance gap increases with $J$, demonstrating its ability to improve generalization and enhance agents' coordination.
Moreover, since \textit{GE-VDAC} utilizes the dimension-reduced embedded features, it has lower complexity than \textit{IE-VDAC} and takes less training time, and the time gap is larger as $J$ increases.

\begin{figure}
  \centering
  \includegraphics[width=.48\textwidth]{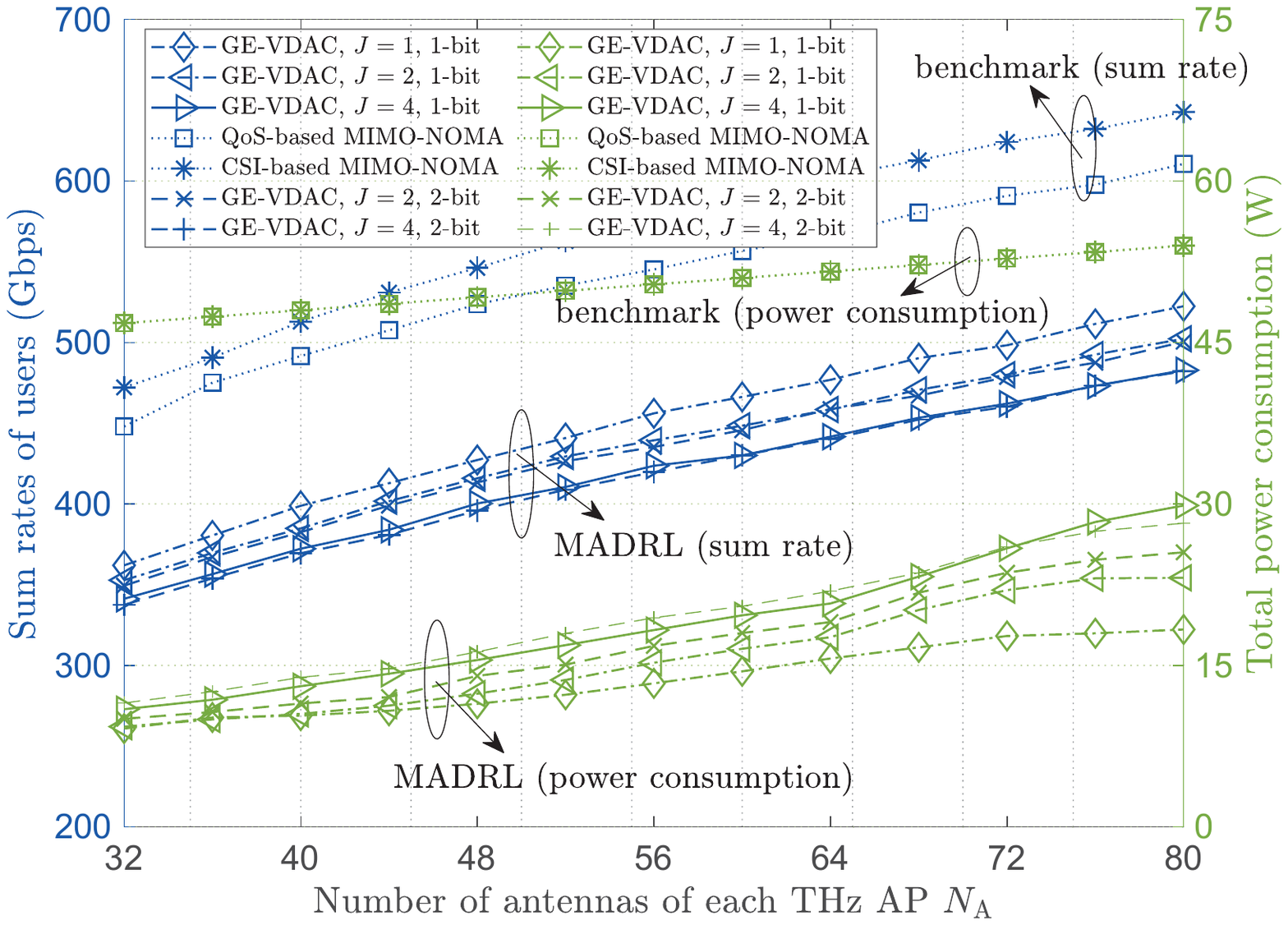}
  \caption{The relationship between sum rate and power allocation.}\label{RP_NA}
\end{figure}

In Fig. \ref{fig_K}, we show the energy efficiency and reliability performance of the smart reconfigurable THz MIMO-NOMA networks under different associated user numbers $K$.
Here, we further consider conventional THz massive MIMO-NOMA without the aid of RISs as benchmarks, which utilizes equal power allocation under channel-based/QoS-based user clustering and NOMA decoding.
We can see that the CSI-based MIMO-NOMA scheme has the lowest reliability of SE users, and the performance gap dramatically grows as the associated users increase.
Without the aid of RISs, the QoS-based MIMO-NOMA scheme can provide higher reliability than CSI-based MIMO-NOMA scheme, but leads to the lowest energy efficiency.
However, the proposed algorithms can achieve the highest global energy efficiency, while guaranteeing the reliability of SE users.
Even when the number of IoT users increases, the reliability of SE users can be guaranteed to be larger than $0.94$, which verifies that the proposed reconfigurable THz MIMO-NOMA can provide on-demand QoS for heterogeneous users with efficient spectral utilization.

\begin{figure}[htbp]
\centering
\subfloat[Energy efficiency.]{\label{EE_NA}
    \includegraphics[width=.48\textwidth]{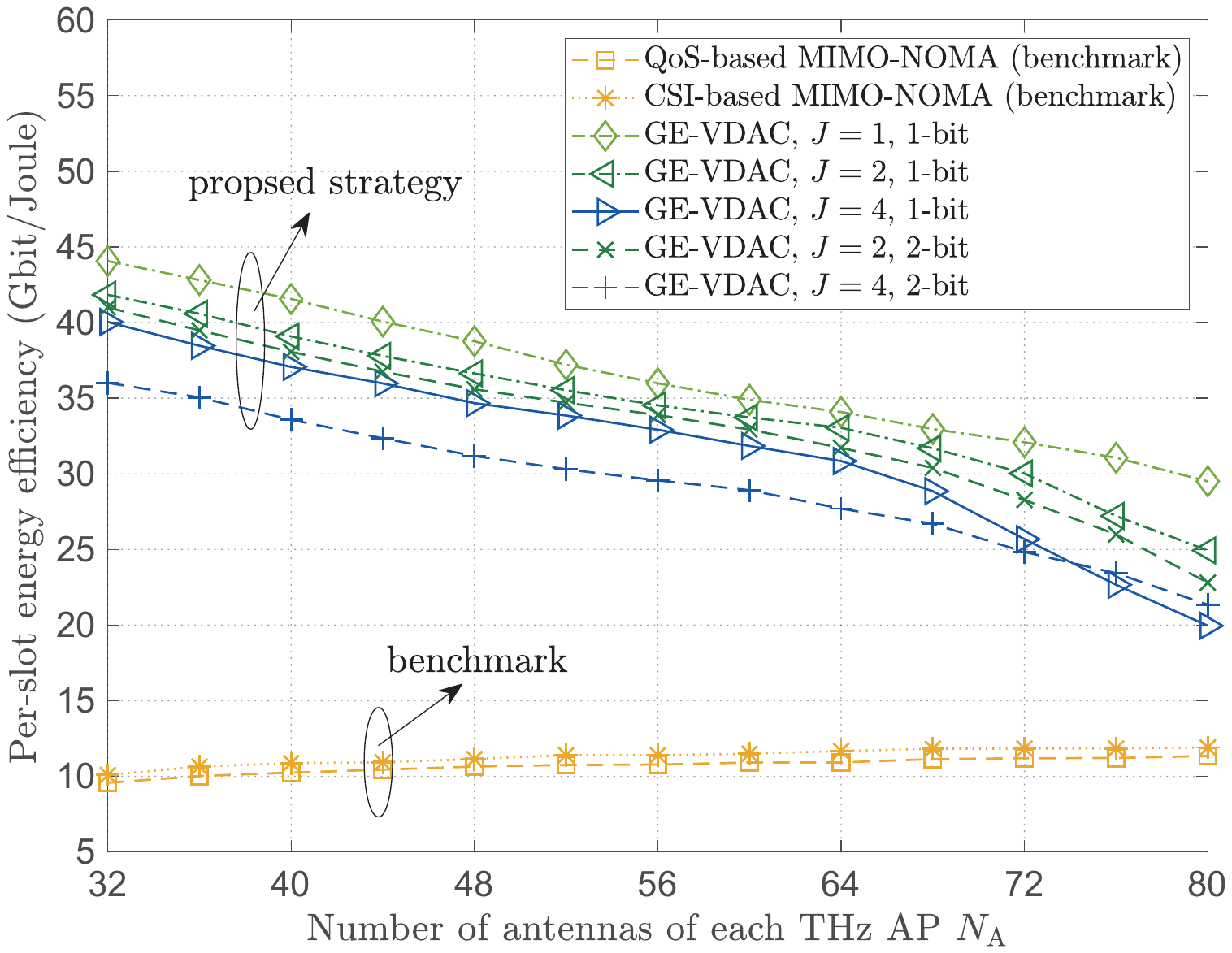}
}
\\
\subfloat[Reliability of SE users.]{\label{Relia_NA}
    \includegraphics[width=.48\textwidth]{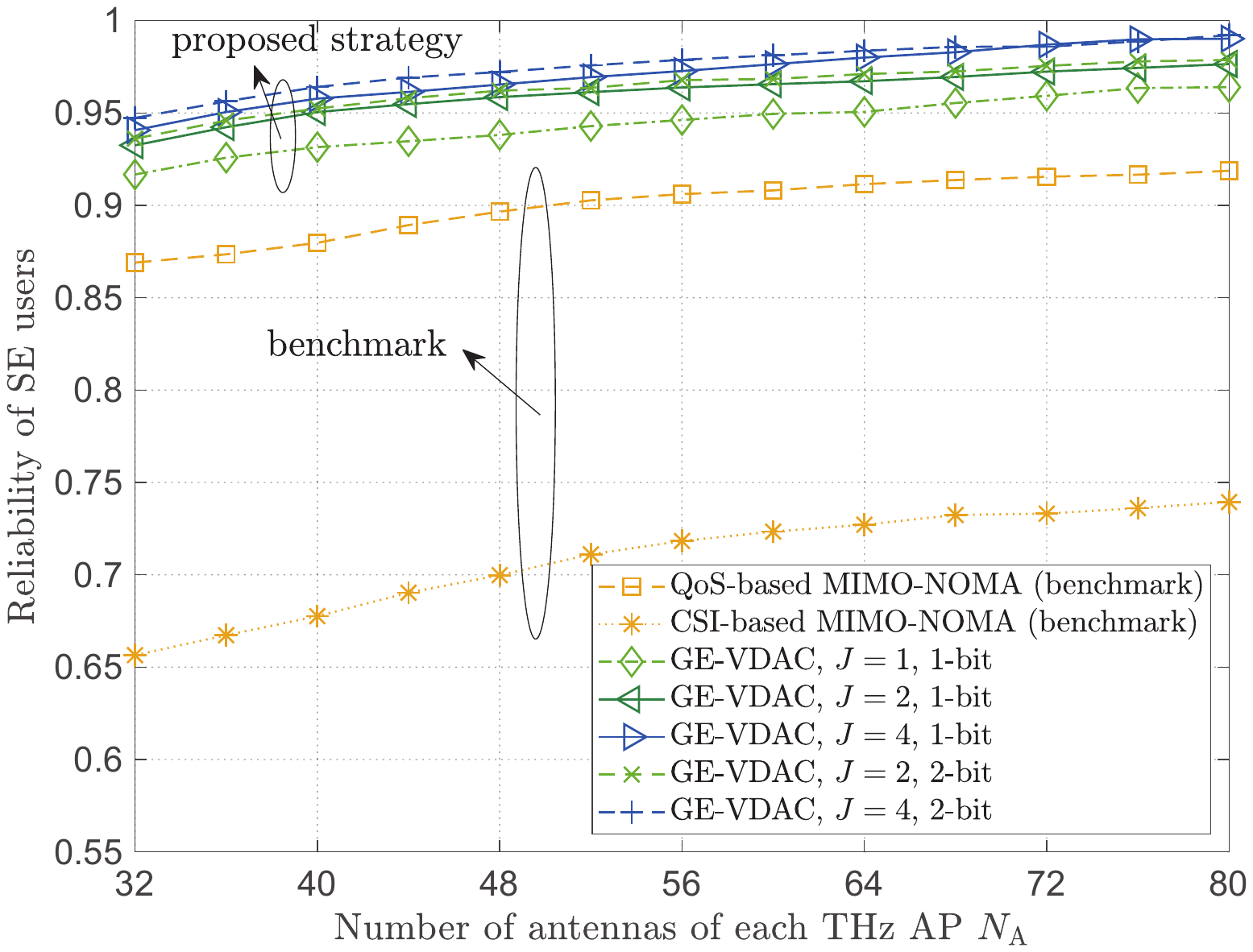}
}
\caption{System performance under different numbers of AP antennas.}
\label{fig_NA}
\end{figure}

Fig. \ref{RP_NA} and Fig. \ref{fig_NA} present the performance of the proposed networks under different number of RISs $N_\mathrm{A}$ and RIS quantization bit $b$.
As shown in Fig. \ref{RP_NA}, when $N_\mathrm{A}$ and $J$ increase, the power consumption increases due to the higher hardware power dissipation.
Meanwhile, the sum rate of users increases with $N_\mathrm{A}$ but decreases with $J$, since the spatial stream interferences become more severe as $J$ growing.
In Fig. \ref{fig_NA}(a), the energy efficiency declines with $N_\mathrm{A}$ and $J$ based on the proposed algorithms due to the higher circuit power dissipation.
Compared with $1$-bit RISs, $2$-bit RISs result in higher power dissipation to achieve competitive data rate.
Nevertheless, $2$-bit RISs can achieve higher energy efficiency when $J=4$ and the number of antennas is large, since they have higher flexibility to suppress the severe spatial interference.
Furthermore, Fig. \ref{fig_NA}(b) shows the reliability of SE users increases with $N_\mathrm{A}$, $J$, and $b$, which demonstrates the effectiveness of massive MIMO structure and the cooperative RISs to avoid blockage.

\section{Conclusion}
In this work, we propose a novel smart reconfigurable THz MIMO-NOMA framework, which empowers customizable and intelligent indoor communications. We first propose specific schemes of customized user clustering, NOMA decoding, and hybrid beamforming. Thereafter, we formulate a long-term Dec-POMDP problem under a joint dynamic RIS element selection, coordinated discrete phase-shift control, and power allocation strategy, which maximizes the network energy efficiency while ensuring heterogenous data rate and reliability for on-demanded users.
To efficiently solve the intractable highly complex MINLP problem, we further develop a decentralized \textit{GE-VDAC} based MADRL algorithm.
We theoretically demonstrate that the \textit{GE-VDAC} algorithm achieves locally optimal convergence and better generalization.
Numerical results show that the proposed algorithm can achieve higher energy efficiency and faster learning speed compared to traditional MADRL methods.
Furthermore, we can also obtain highly customized communications through efficient resource utilization.


%

\appendices

\section{Proof of Lemma \ref{Lemma_LyapunovDrift}} \label{Proof_LyapunovDrift}
Considering the definition of $q_{nk}^{m}$ and $\left(\left[x\right]^{+}\right)^{2}\leq x^{2}$, we have
\begin{equation}\label{q_2_bound}
\begin{split}
&\frac{1}{2}\left(q_{nk}^{m}(t+1)\right)^{2}
\leq \frac{1}{2}\left[q_{nk}^{m}(t)-R_{nk}^{m}(t)+A_{nk}^{m}(t)\right]^{2} \\&
\leq \frac{1}{2}\left(q_{nk}^{m}(t)\right)^{2}+C_{nk}^{m,\mathrm{q}}+q_{nk}^{m}(t)\left(A_{nk}^{m}(t)-R_{nk}^{m}(t)\right).
\end{split}
\end{equation}

Similarly, considering the definition of $Z_{nk}^{m}$, we have
\begin{equation}\label{z_2_bound}
\begin{split}
&  \frac{1}{2}\left(Y_{nk}^{m}(t+1)^{2} - Y_{nk}^{m}(t)^{2} \right) \\&
\overset{(a)}{\leq} \frac{1}{2}\left(q_{nk}^{m}(t+1)\right)^{2}+C_{nk}^{m,\mathrm{Y}}+Y_{nk}^{m}(t)q_{nk}^{m}(t+1),
\end{split}
\end{equation}
where $(a)$ is due to $q_{nk}^{m,\max}\epsilon_{nk}^{m}$ and $q_{nk}^{m}(t)$ are non-negative.
Substituting \eqref{q_2_bound} into \eqref{z_2_bound} and considering the fact that $q_{nk}^{m}(t+1) = q_{nk}^{m}(t)-R_{nk}^{m}(t)+A_{nk}^{m}(t)$ due to the nonempty traffic queue, we can obtain
\begin{equation}\label{z_2_bound2}
\begin{split}
& \frac{1}{2}\left(Y_{nk}^{m}(t+1)^{2} - Y_{nk}^{m}(t)^{2} \right) \\&
\leq
C_{nk}^{m,\mathrm{q}}+C_{nk}^{m,\mathrm{Y}}
-\left(q_{nk}^{m}(t)+Y_{nk}^{m}(t)\right)R_{nk}^{m}(t)+ B_{nk}^{m}(t)
.
\end{split}
\end{equation}

Substituting \eqref{q_2_bound} and \eqref{z_2_bound2} into $\Delta \mathcal{L}_{nk}^{m}(t)=\frac{1}{2}\big[q_{nk}^{m}(t+1)^{2}-q_{nk}^{m}(t)^{2}\big]+\frac{1}{2}\big[Y_{nk}^{m}(t+1)^{2}-Y_{nk}^{m}(t)^{2}\big]$ concludes the proof.

\section{Proof of the Theorem \ref{Convergence}} \label{Proof_Convergence}
For the sake of expression, we ignore the time slot index $t$ in the following proof.
The joint policy can be written as the product of the graph embedding sub-policy and the independent action generation sub-policies:
\begin{equation}\label{joint_policy}
\pi\left(\mathbf{a}|\mathbf{s}\right)
=\prod_{u}\prod_{i}\pi_{\theta_{\mathrm{G}}}\left(\widetilde{\mathbf{\mathbf{z}}}_{i}^{u}|\mathbf{O},\mathbf{D}\right)
\pi_{\theta_{\mathrm{A}}}^{u}\left(\mathbf{a}_{i}^{u}|\widetilde{\mathbf{\mathbf{z}}}_{i}^{u}\right).
\end{equation}

Therefore, the policy gradient in \eqref{policy_gradient} can be rewritten as
\begin{equation}\label{pg_2}\small
\begin{split}
\mathbf{g}_{\pi}
& =\mathbb{E}_{\pi}\left[\sum_{u}\sum_{i}\nabla_{\bm{\theta}_{\pi}}\log\left(\pi_{\theta_{\mathrm{G}}}\left(\widetilde{\mathbf{\mathbf{z}}}_{i}^{u}|\mathbf{O},\mathbf{D}\right)
\pi_{\theta_{\mathrm{A}}}^{u}\left(\mathbf{a}_{i}^{u}|\widetilde{\mathbf{\mathbf{z}}}_{i}^{u}\right)\right)A\left(\mathbf{a},\mathbf{s}\right)\right]
\\&
=\mathbb{E}_{\pi}\left[\nabla_{\bm{\theta}_{\pi}}\log
\left(\prod_{u}\prod_{i}\pi_{\theta_{\mathrm{G}}}\left(\widetilde{\mathbf{\mathbf{z}}}_{i}^{u}|\mathbf{O},\mathbf{D}\right)\pi_{\theta_{\mathrm{A}}}^{u}\left(\mathbf{a}_{i}^{u}|\widetilde{\mathbf{\mathbf{z}}}_{i}^{u}\right)\right)
A\left(\mathbf{a},\mathbf{s}\right)\right]
\\& \normalsize
\overset{\eqref{joint_policy}}{=} \mathbb{E}_{\pi}\left[\nabla_{\bm{\theta}_{\pi}}\log\pi\left(\mathbf{a}|\mathbf{s}\right)\left(Q\left(\mathbf{a},\mathbf{s}\right)-V_{\mathrm{tot}}\left(\mathbf{s}\right)\right)\right],
\end{split}
\end{equation}
where $\bm{\theta}_{\pi}=\left[\left\{ \mathbf{\bm{\theta}}_{\mathrm{G}}^{u}\right\} ,\left\{ \mathbf{\bm{\theta}}_{\mathrm{A}}^{u}\right\} \right]$ is the joint policy parameter vector.

Denoting $d^{\pi(s)}$ as the discounted ergodic state distribution, we can obtain \cite{VDAC_Su_2021}, \cite{PG_Sutton_2000}
\begin{equation}\label{baseline}
\begin{split}
&\nabla_{\bm{\theta}_{\pi}}\log\pi\left(\mathbf{a}|\mathbf{s}\right)V_{\mathrm{tot}}\left(\mathbf{s}\right)
\\
=&\sum_{\mathbf{s}}d^{\pi(s)}V_{\mathrm{tot}}\left(\mathbf{s}\right)\nabla_{\bm{\theta}_{\pi}}\sum_{\mathbf{a}}\log\pi\left(\mathbf{a}|\mathbf{s}\right)
\\
=&\sum_{\mathbf{s}}d^{\pi(s)}V_{\mathrm{tot}}\left(\mathbf{s}\right)\nabla_{\bm{\theta}_{\pi}}1
=\mathbf{0}.
\end{split}
\end{equation}

Substituting \eqref{baseline} into \eqref{pg_2}, we have
\begin{equation}
\mathbf{g}_{\pi}
=\mathbb{E}_{\pi}\left[\nabla_{\bm{\theta}_{\pi}}\log\pi\left(\mathbf{a}|\mathbf{s}\right)Q\left(\mathbf{s},\mathbf{a}\right)\right],
\end{equation}
which leads to a standard policy gradient for single-agent actor-critic algorithm.
In \cite{AC_Konda_2000}, it is proven that an actor-critic based on such gradient can converge to a local maximal expected return, which satisfies \eqref{optimum}.
This ends the proof.


\ifCLASSOPTIONcaptionsoff
  \newpage
\fi

\end{document}